# Scalable orthogonal delay-division multiplexed OEO artificial neural network trained for TI-ADC equalization

ANDREA ZAZZI,[1] ARKA DIPTA DAS,[1] LUKAS HÜSSEN,[2] RENATO NEGRA,[2] AND JEREMY WITZENS[1,*]

[1]*Institute of Integrated Photonics, RWTH Aachen University, Campus Blvd. 73, 52074 Aachen, Germany*
[2]*Chair of High-Frequency Electronics, RWTH Aachen University, Kopernikusstraße 16, 52074 Aachen, Germany*
*\*jwitzens@iph.rwth-aachen.de*

**Abstract:** We propose a new signaling scheme for on-chip optical-electrical-optical artificial neural networks that utilizes orthogonal delay-division multiplexing and pilot-tone based self-homodyne detection. This scheme offers a more efficient scaling of the optical power budget with increasing network complexity. Our simulations, based on a 220 nm SOI silicon photonics technology, suggest that the network can support $31 \times 31$ neurons, with 961 links and freely programmable weights, using a single 500 mW optical comb and an SNR of 21.3 dB per neuron. Moreover, it features a low sensitivity to temperature fluctuations, ensuring that it can be operated outside of a laboratory environment. We demonstrate the network's effectiveness in nonlinear equalization tasks by training it to equalize a time-interleaved ADC architecture, achieving an ENOB over 4 over the entire 75 GHz ADC bandwidth. We anticipate that this network architecture will enable broadband and low latency nonlinear signal processing in practical settings such as ultra-broadband data converters and real-time control systems.

## 1. Introduction

Mixed optical-electrical signal processing is utilized for broadband and low latency operation, and is applied in various fields including artificial neural networks (ANNs),[1-8] high-speed analog-to-digital (ADC)[9-15] and digital-to-analog (DAC) converters,[16,17] and phased-array based imaging.[18] Optical-electrical-optical (OEO) ANNs use optical signal processing for weighting and summation, while electrical signal regeneration is employed between layers. Several opto-electronic methods have also been developed for implementing the nonlinear activation function, such as using ring modulators[6,7] or feed-forward electro-optic processing.[4] Neuromorphic hardware accelerators have been used for convolutional processing applied to digit recognition,[5] end-to-end four-class classification of handwritten letters,[7] and equalization of fiber nonlinearities in long-haul fiber optic communications.[6] However, the number of independent weights implemented on chip remains modest, with up to four artificial neurons (ANs) per layer[1,6,7] or up to four independent weights in a convolutional network.[5] The challenge of scalability is an outstanding issue, in contrast to ANNs that rely on free-space optics and parallelized electro-optic processing with high pixel-count spatial phase modulators.[2] These are, however, much bulkier setups with limited operation speed.

On-chip solutions have relied on coherent signal processing[1] or on signal transduction and weighting with resonant ring-based devices.[6,7] Both approaches are very sensitive to phase errors, and resonant devices, in particular, require precise stabilization against temperature fluctuations and fabrication variability.[19] The required phase shifters (PSs) present challenges to scaling due to factors such as power consumption in case of thermally actuated devices,[20] the excess optical losses of capacitively switched carrier accumulation devices,[21] the requirement for stable and reversible grey scale programing of switchable remanent materials,[22,23] or the size and drive voltage requirements of microelectromechanical (MEMS)[24] and stress-optic devices.[25] Additionally, operating with increasing AN counts presents significant challenges for the control system. Therefore, it is crucial to engineer systems with a high tolerance to phase errors and environmental temperature fluctuations.

The photonic crossbar architecture shown by Feldmann *et al.*[5] addresses some scalability issues, as summations are implemented by incoherent superposition of signals with different carrier frequencies and weighting is implemented with non-resonant devices consisting simply of a phase change material with programmable absorption. However, the implemented 2×2 convolution kernel, while being applied in parallel to large datasets, remains of modest size and corresponds to a 4-by-4 arbitrary matrix multiplication. The non-selective broadband cross-connects create excess insertion losses (ILs) that accumulate as the crossbar array is scaled up. Additionally, precise wavelength division multiplexers with narrowly spaced channels are required, and these have been implemented off chip. Although integrating such devices into the silicon photonics platform is more practical with silicon nitride (SiN) waveguides,[26] obtaining the required untuned performance remains challenging[27] due to the free spectral range (FSR) of semiconductor laser pumped combs.[28,29] Nevertheless, reliance on incoherent processing and the distribution of information across optical combs shows great promise for increasing network scalability.

While many OEO architectures are modeled on wavelength division multiplexed (WDM) networks,[6-8] that are very effective for long-haul communications, on-chip ANNs present a very different set of constraints. To scale up on-chip ANNs, innovative signal multiplexing schemes are needed that effectively use available light and are not limited by resonant devices or other thermally sensitive optical filters. Building on the concepts of incoherent signal summation and spectrally broadband signal processing, we introduce a new on-chip optical network architecture that allows complete interconnection between two layers of ($N \times$) ANs, with configurable optical signal weighting and summation, without any resonant device, with near athermal operation, and improved optical power efficiency. Our concrete device designs and network modeling suggest that this architecture can scale up to $31 \times 31$ ANs, with a total of 961 logical interconnects and independent weights, using a 500 mW optical power level that can be injected into a single-mode SiN waveguide without reaching its damage threshold.[30,31]

The proposed architecture distributes the information of a single logical channel, defined as the information broadcasted by an upstream AN, over the entire spectrum of an optical comb, while maintaining orthogonality between channels. This approach uses self-referenced, pilot-tone based homodyne detection to enable channel selection and summation, which halves the effect of modulator ILs without incurring the circuit complexity and excess power consumption usually associated with coherent detection. The pilot tone power can be dynamically allocated at downstream ANs, which improves the scaling of received signal strengths to $\mathcal{O}(1/N^{1.5}M^{0.5})$, where $M$ is the effective number of channels received by a given AN with significant weights. This scaling outperforms that of arbitrarily configurable WDM networks,[6] in which the required power scales as $\mathcal{O}(1/N^2)$ regardless of the sparsity of the dynamically programmed matrix operation. The architecture is based on feed-forward, balanced optical paths, which makes it thermally stable and easier to control. This architecture thus presents two significant improvements over the state-of-the-art that improves its scalability: (i) A better allocation of optical power to utilized on-chip data paths when sparse weights are utilized, resulting in a better signal-to-noise ratio (SNR), and (ii) a high temperature tolerance arising from a filter-less network, making the control system manageable as the system is scaled up. The optical power budget is particularly important for system scalability, since the damage threshold of integrated waveguides limits available optical power. Other aspects are inherited from other OEO network configurations. In particular, our system also presents the potential for very low latency, broadband operation, but suffers from the electrical power consumption overhead associated to repeated transduction between the electrical and optical domains.

We introduce the high-level architecture concept (Section 2) and its configurability (Section 3), analyze its practical implementation in presence of noise, dispersion and device nonidealities (Section 4), and evaluate its applicability to the nonlinear equalization of a time-interleaved optically enabled ADC architecture featured as an exemplary use-case (Section 5). The convergence of a training algorithm applied directly to the PS settings is verified and the system-level signal integrity evaluated.

## 2. Network architecture

At the first level of analysis, the multiplexing scheme is based on delayed versions of an incoherent carrier, which do not interfere with each other when differential delays exceed its coherence time. A broadband light source emits light in a spectrum $\Delta \nu$ with a small coherence length $L_c \approx c_0/\Delta \nu n_g$, where $c_0$ is the speed of light in vacuum and $n_g$ the group index of the waveguide in which it is propagating. Figure 1(a) shows a simplified schematic of the network. The light is first split in a reference branch, later serving as a pilot tone, and $N$ additional branches, each supplying light to an AN of the upstream layer, which phase modulates it according to the signal it is broadcasting. The modulated optical signals are then fed through an $N \times N$ network that distributes each of its inputs equally onto each of its outputs, such as a multi-mode interferometer (MMI) or a star coupler.[32] Each of the resulting signals are then sent to an AN of the downstream layer. The pilot tone is also distributed to the downstream ANs, which use it to demodulate the incoming signals using an interferometer.

The use of delay loops in the waveguides prior to the distribution network allows for the distinguishing of signals from each other even after they have been superposed, provided length increments exceed the coherence length of the light. Signals then simply sum up in power without interference if they are lowpass filtered / integrated by a sufficient amount (see Appendix I). At the downstream ANs, only signals delayed by the same amount as the pilot tone can interfere with it and create a differential signal at the balanced photodiode (BPD) pair implemented at the demodulator output. For that purpose, the demodulators are provided with differential group delays in their upper and lower branches that synchronize only selected signals with the pilot tone. The resulting interference signal is recorded by the BPD, reamplified by a transimpedance amplifier (TIA) provided with a nonlinear transfer function implementing the activation function, and finally used to drive the high-speed phase modulators providing optical signals to the next layer.

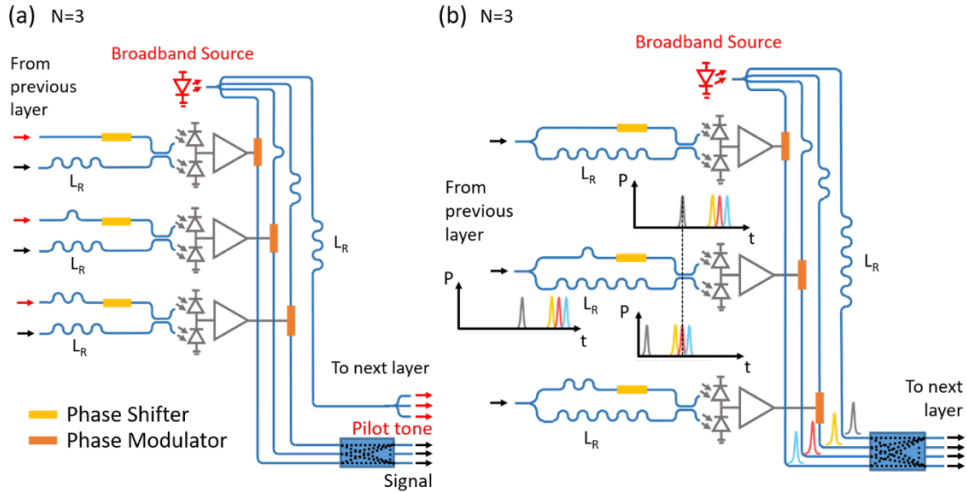

Fig. 1. Orthogonal delay-division multiplexing scheme represented for $N = 3$. (a) Power-budget-optimized network architecture, with pilot tone distributed to the downstream layer with separate waveguides. Red and black arrows represent the pilot tone and signals, respectively. (b) Fabrication-tolerant architecture in which the pilot tone and signals are guided together. A time-domain representation of a pulse-based network is also shown.

To simplify network routing and minimize the length over which signals and pilot tone propagate independently, for best relative phase stability, the distribution network can be extended to $(N+1) \times (N+1)$ ports to broadcast the pilot tone and modulated signals through the same interconnect waveguides, Fig. 1(b). This results in a 3-dB demodulated signal penalty at the downstream ANs and increases the required spectral width of the light source, as explained below, but greatly enhances network robustness. Signals can then be routed between AN layers on different chips with optical fibers without compromising the relative phase stability.

In the transmitter (Tx) network before the distribution network, signals are delayed by $nL_0$, with $n = 0 \dots N-1$ indexing the AN and $L_0 \geq L_c$. The pilot tone is delayed by a different amount, $L_R$, that can be as small as $NL_0$ in Fig. 1(a) but needs to be at least $(2N-1)L_0$ in Fig. 1(b): For the demodulators in Fig. 1(b) to resynchronize signals and pilot tone, the signals need to travel through the lower branch and the pilot tone through the upper branch. However, half the signals' power also travels through the upper branch, resulting in cumulative delays between 0 and $(2N-2)L_0$ (i), and half the pilot tone power travels through the lower branch, resulting in a cumulative delay of $2L_R$ (ii). Besides the 3-dB excess loss, this does not result in further penalties as long as these cannot interfere with each other or any of the other delayed beams. Since the pilot tone sees cumulative delays between $L_R$ and $L_R + (N-1)L_0$ after travelling through the upper branch, (i) results in the conditions $[L_R, \dots, L_R + (N-1)L_0] \cap [0, L_0, \dots, (2N-2)L_0] = \emptyset$ and thus $L_R \geq (2N-1)L_0$.

The broadband light source can be implemented as a low noise comb. A simple intuition can be gained in the case of pulsed operation, as illustrated in Fig. 1(b). Each of the upstream ANs then phase modulates a pulse that occupies a different time slot in a time window given by the repetition time of the comb, $T_{rep} = 1/\delta\nu$, with $\delta\nu$ its FSR. Since the pulse width is in the order of $1/\Delta\nu$, the delay length increment $L_0$ needs to be at least $c_0/n_g\Delta\nu$, which is the criterium already introduced above. The demodulator generates a differential photocurrent only when a signal pulse is synchronized in time with a pilot tone pulse, so that it can interfere with it. It should be noted that while the information is encoded in dedicated time slots, this is not a time-multiplexed communication system in the conventional sense, as the electronics of the ANs record the signal over the entire comb repetition time $T_{rep}$, or a multiple thereof, and the channel selection is achieved by sampling with the pilot tone pulse. Importantly, the network can operate irrespectively of whether the comb source emits pulses or has a dispersed output. In the latter case, the time domain picture is no longer helpful to guide the intuition. However, orthogonality between the logical channels and selective demodulation are maintained, as derived in Appendix I and numerically shown in Section 4. Operation of the network with dispersed pulses is an essential feature, as it reduces the peak power at the chip interface and in on-chip waveguides for a given network size, and thus significantly increases the ANN size that can be achieved without damage.

The two scenarios described above are special cases of a common theoretical framework, in which information is distributed over the entire light spectrum in an orthogonal manner, to which we refer to as orthogonal delay-division multiplexing (ODDM). The information can be retrieved by applying an inverse (I-) discrete Fourier transform (DFT) to the set of complex-valued comb line amplitudes (Appendix I). The number of slots available for logical channels, including unused ones and the slot taken up by the pilot tone, are equal to the number of comb lines in a square-shaped spectrum. These slots are further referred to as ODDM channels to distinguish them from logical channels, the subset that carries data. Since the maximum delay occurring in Fig. 1(b) is $2L_R \geq (4N-2)L_0$, at least $4N-1$ comb lines are required. This is about two times more than for the architecture shown in Fig. 1(a). However, the resulting reduction in spectral efficiency is a secondary concern for the on-chip networks conceived here, provided sufficiently broadband light sources and on-chip devices are available.

We give a first estimate of the received signal strengths here, with a more complete model derived in Section 3 for an AN demodulating several incoming signals. The pilot tone's amplitude in the upper branch is denoted as $E_R$, and the amplitude of the modulated optical signal corresponding to channel *m* in the lower branch as $E_m$. Additionally, the phase applied by the upstream AN of index *m* is represented as $\varphi_m$, and the responsivity of the PDs as $R$. The differential photocurrent received by the TIA can be calculated as $I_p = 2RE_RE_m\sin(\varphi_m)$. Allocating half the available power to the pilot tone and the other half to the modulators in the initial splitter network maximizes this expression. Downstream ANs experience optical signal strengths scaling as $E_m \propto 1/N$, whereas the pilot tone strength reduces only as $E_R \propto 1/\sqrt{N}$ when allocated to demodulating a single channel, as in Fig. 1. As a result, $I_p$ scales as $\mathcal{O}(1/N\sqrt{N})$. More precisely, it can be expressed as

$$I_p = 10^{-\frac{IL_R}{10} - \frac{IL_{mod}}{20}} \frac{RP_c}{2N\sqrt{N}} \sin(\varphi_m + \gamma_{pm} - \eta_{pm}) \tag{1}$$

where $IL_R$ is the excess IL seen by both pilot tone and optical signals and $IL_{mod}$ represents the excess modulator and splitter network losses applied only to the signals. The distribution network introduces a phase $\gamma_{pm}$ that can be compensated by applying an additional phase $\eta_{pm}$ in the upper demodulator branch using a PS. Here, *m* and *p* are the indices of the upstream and downstream ANs, respectively, and $P_c$ corresponds to the power emitted by the light source.

The inherent temperature stability of the network can already be inferred from Fig. 1(b): While temperature fluctuations result in different phase errors being applied to the different Tx signal branches as a consequence of the different delay lengths, for each demodulated signal the pilot tone travels through delays of corresponding length in the demodulator (with the converse applying to the modulated signal relative to the Tx pilot tone delay), such that the phase errors cancel out as part of the homodyne detection. This is the same concept as resulting in the high temperature tolerance of balanced Mach-Zehnder modulators (MZM). In an MZM, only the relative phase between the two branches matters, as opposed to e.g. micro-ring modulators for which the absolute phase is the relevant metric. If the two arms of the MZM are perfectly balanced and subjected to the same temperature swing, the applied phase increments in the two branches will be identical and the bias point of the MZM remains unchanged by design. Any phase walk-offs occur only due to device implementation mismatches, which will be further discussed in Section 4 for the present network.

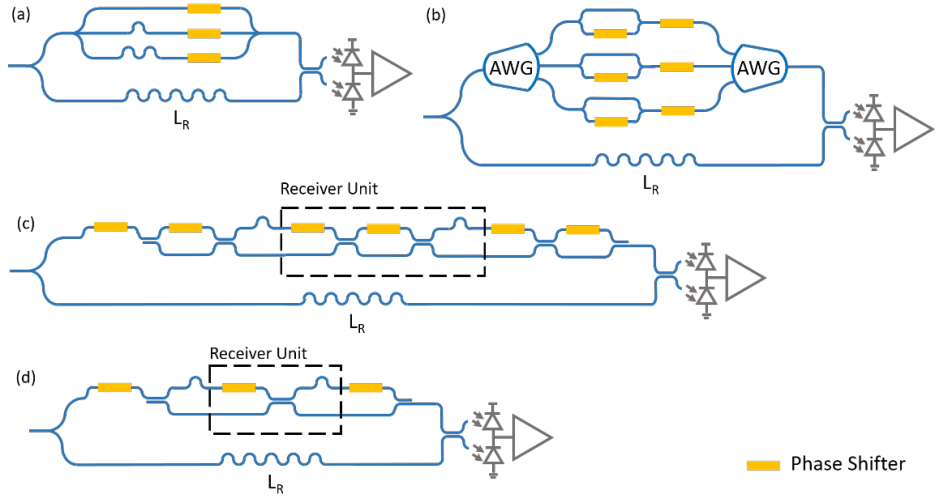

Fig. 2. Signal weighting schemes. In (a) phases can be set independently in the top demodulator branch at the expense of reducing the amplitude proportionally to the number of paths. In (b) this is alleviated by introducing arrayed waveguide gratings to (de-)multiplex the comb lines, allowing, in principle, lossless signal recombination. However, AWGs require substantial chip area. Significant reduction of the demodulator size is achieved by simultaneously allowing the pilot tone to cross each path in the upper demodulator branch of (c), so that a programmable superposition of all possible combinations of delays is obtained. (d) Further simplified network in which the embedded MZIs have been replaced by static splitters with an off-3-dB splitting ratio.

## 3. Arbitrary signal weighting

The network described so far is restricted to the implementation of point-to-point logical links and does not yet allow an AN to receive signals from multiple upstream neurons, making it yet unsuitable for the implementation of an ANN.

To address this limitation, the demodulator can be replaced by the device shown in Fig. 2(a). It generates a superposition of delayed pilot tones in its upper branch, resulting in a summation of corresponding demodulated signals at the BPD. The phases $\eta_{pm}$ associated with each of the $M$ possible delay paths in the upper branch are set by PSs and can be used to compensate the phases $\gamma_{pm}$ accumulated in the distribution network and to determine the small-signal weights for the corresponding logical channels, even without additional programmability. However, this architecture also introduces a drawback, namely a $1/M$ amplitude reduction for each of the delayed versions of the pilot tone, thereby reducing the received differential current strength by the same amount. The $1/\sqrt{M}$ amplitude reduction at the combiner, in particular, is unnecessary.

Conceptually, this issue can be improved by the architecture shown in Fig. 2(b). The one-by-$M$ splitters are replaced by 1-by-$Q$ arrayed waveguide gratings (AWGs), with $Q$ the number of comb lines, so that the light can be theoretically split and recombined without losses. In between the AWGs, the comb line phase and amplitude are modified, allowing an initial pilot tone with complex-valued comb line amplitudes $v_{q,in}^R$, $q \in [0 \ldots Q-1]$, to be transformed into an arbitrary superposition of delayed pilot tones given by coefficients $v_{q,out}^R$, provided that $|v_{q,out}^R| \leq |v_{q,in}^R|$. For example, assuming $|v_{q,in}^R| = 1$, the pilot tone can be transformed into $v_{q,out}^R = e^{-imq\delta\omega\tau_0}e^{i\eta_{pm}}$, corresponding to a group delay $m\tau_0 = m L_0 n_g/c_0$ and a phase delay $\eta_{pm}$, with $\delta\omega = 2\pi\delta\nu$ and $\tau_0$ the group delay increment induced by the delay length $L_0$. This allows the reception of logical channel $m$ with the full strength of the pilot tone. More generally, the summation over logical channels

$$I_p = 10^{-\frac{IL_R}{10}-\frac{IL_{mod}}{20}} \frac{RP_c}{2N\sqrt{N}} \sum_{m=0}^{N-1} |\tilde{v}_m^R| \sin(\varphi_m + \gamma_{pm} - \eta_{pm}) \qquad (2)$$

is obtained, wherein the weight coefficients $\widetilde{\boldsymbol{v}}^{\boldsymbol{R}} = \left[|\tilde{v}_m^R|e^{i\eta_{pm}}\right]_m$ are obtained by applying an IDFT to the set of reference comb line amplitudes $\boldsymbol{v}_{\boldsymbol{out}}^{\boldsymbol{R}} = \left[v_{q,out}^R\right]_q$ (see Appendix I).

The constraint $|v_{q,out}^R/v_{q,in}^R| \leq 1$ bounds the obtainable weights. In combination with the Parseval theorem, this results in $\sum_m |\tilde{v}_m^R|^2$ to be bounded by $\sum_q |v_{q,in}^R|^2$, the incoming pilot tone power. If $M$ signals are jointly received, the coefficients $|\tilde{v}_m^R|$ thus need to scale down by an average $\sim 1/\sqrt{M}$ (a factor $\sqrt{M}$ better than in Fig. 2(a)). Equivalently, the power of the pilot tone is thus dynamically allocated to $M$ demodulated logical channels, according to the PS configurations, instead of being statically split over a fixed number of channels and recombined with excess losses as in Fig. 2(a), improving the scaling of the network.

Since the circuit shown in Fig. 2(b) uses $2N$ degrees of freedom to generate $N$ logical weights, we have some level of redundancy, that allows us to maximize the demodulated signal strengths achievable overall. Indeed, we can null the coefficients $\gamma_{pm}$ in Eq. (2), arising from the distribution network, by applying an equal phase $\eta_{pm}$ to the corresponding pilot tone component $\tilde{v}_m^R$. This maximizes the derivative of $I_p$ relative to $\varphi_m$ and thus the achievable signal strength. The weight is then set by means of $|\tilde{v}_m^R|$. This minimizes the amplitude $|\tilde{v}_m^R|$ required for a given signal strength and frees up pilot line power that can be allocated to other channels. A non-zero $\gamma_{pm} - \eta_{pm}$, on the other hand, plays the equivalent role of a bias applied to the input signal $\varphi_m$. Due to the nonlinearity of the sine curves, such biases are individual for each input signal, as opposed to the configuration of a conventional ANN. They bias Eq. (2)

away from the quadrature point, effectively also modifying the activation function of the neuron for that particular input.

While this results in adequate functionality, the corresponding photonic circuit is too complex to be scalable. In particular, each AWG requires delay lines ranging from $L_0$ to $(Q-1)L_0$, $Q \geq (4N-1)$, resulting in a cumulative length of $\sim 16N^2 L_0$ per AN and a prohibitive amount of chip real estate that scales, for the whole ANN, as the cube of the channel count. Instead, we envision the much simpler network shown in Fig. 2(c), that requires a cumulative delay length $(N-1)L_0$ in its upper branch, the theoretical minimum required to access all the channels. Its upper branch is formed by $N$ identical segments labeled as receiver units. These consist in 2-by-2 networks implementing a unitary transform determined by two PSs separated by 3-dB directional coupler splitters (DCS),[33,34] after which the light can either follow the lower waveguide and be directly transmitted, or the upper waveguide in which it is delayed by $L_0$. In total, the same number ($2N$) of degrees of freedom is provided. However, a challenge resides in the mapping between the PS settings and the weights that is not one-to-one. Rather, the PS settings have to be trained by applying the back-propagation algorithm directly to them (Section 5).

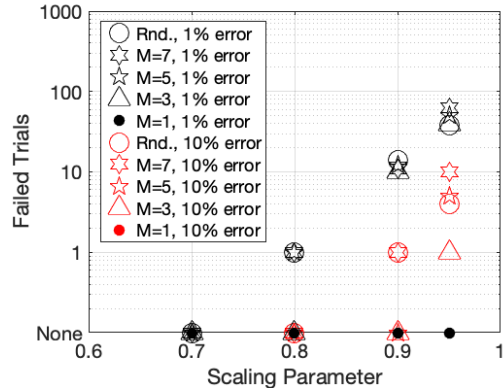

Fig. 3. Numerical trials for the validation of the weight space addressable by the demodulator architecture shown in Fig. 2(c). Trials are categorized as having failed when a suitable PS-configuration cannot be found to reach the randomly generated set of target weights. For each data point, corresponding to a given weight scenario and scaling parameter $\mu$, 1000 random trials are run and the failures with actual-to-target deviations above 1% and 10% counted.

Extensive numerical modeling indicates that the addressable parameter space is close to that of Fig. 2(b): Numerical trials consisting in searching the correct PS settings for a targeted weight combination are summarized in Fig. 3 for a demodulator receiving up to 7 logical channels. Each data point consists in a thousand numerical trials. Several scenarios consist in randomly generating all 7 weights ("Rnd.") or setting a random subset of $M$ weights to be equal to each other but each correcting a random phase $\gamma_{pm}$, in which case independent trials were made for $M = 1, 3, 5$ and $7$. While the tunable delay line implementation shown in Fig. 2(b) can reach any point of the weight-space provided $|v^R_{q,out}/v^R_{q,in}| \leq 1$, for the evaluation of Fig. 2(c) we introduce a scaling factor $\mu \leq 1$ with the coefficients $v^R_{q,out}$ rescaled such that $max_q(|v^R_{q,out}/v^R_{q,in}|) = \mu$. The smaller $\mu$, the more reliably suitable PS coefficients can be found, but the smaller also the overall weights that are programmed. All 5000 trials were successful for $\mu = 0.7$. At $\mu = 0.8$, all but 3 out of 5000 trials were successful, with a maxim deviation between targeted and obtained weights below 10% for these. For $\mu = 0.9$ and above, an increasing number of trials fails, but even at $\mu = 0.95$ more than 90% succeed for all scenarios. Consequently, the demodulator shown in Fig. 2(c) is almost equivalent to the one in Fig. 2(b), with possibly a small equivalent optical power budget penalty arising from the boundaries of the achievable weight space being pulled in. The deviation between actual and targeted weight vectors was evaluated as the magnitude of the error vector divided by the magnitude of the target vector, calculated as the square root of the L2 norm.

This demodulator architecture offers another essential advantage over the AWG-based implementation due to its insensitivity to correlated process variations during photonic integrated circuit (PIC) fabrication. Unlike AWG passbands, which require absolute process control to maintain alignment with comb line frequencies, the delay lines in the demodulator only need to be matched to those used in the Tx subsystem, making reproducibility across the chip sufficient.

A simplified version of the demodulator in Fig. 2(d) replaces the tunable Mach-Zehnder interferometers (MZI) interposed between the delay line segments by DCSs with a fixed splitting ratio. This simplified architecture is used in the network evaluation described in Section 5, with 75%/25% power splitting. While this removes some degrees of freedom from the network and imposes in particular an unequal dynamic range for achievable weights, which favors input signals with a mid-range delay-index, this was found to be very suitable for typical nonlinear equalization tasks. Such an unequal weight distribution is a-priori expected from a trained equalizer, for which the central taps are closest to target.

## 4. Noise, dispersion and device nonidealities

So far, the ANN architecture has been described in abstract terms. This section describes the practical implementation of a 31-by-31 ANN that takes various factors into account, such as noise, group delay mismatch, dispersion, non-ideal device transfer functions, and realistic comb shapes and electronic transfer functions. To ensure realistic component characteristics, the components were designed into the process of Advanced Micro Foundry (AMF), which supports SiN waveguides in the back-end-of-line (BEOL)[26] in addition to the silicon-on-insulator (SOI) based devices, as well as a back-end-open module compatible with the implementation of silicon-organic-hybrid (SOH) modulators.[35] These modulators are ideal for the architecture modeled here, as they combine a wide optical bandwidth with efficient modulation and are compatible with lumped element driving at the signaling rates investigated below. The low $V_\pi L$ of these modulators allows direct driving with CMOS electronics and their low ILs improve the scalability of the network given the available optical power, as limited by the damage threshold of the waveguide technology used for the splitter network. The damage threshold, in turn, is significantly improved by implementing the splitter network in the SiN layer.[30,31]

Significant progress has been made in the generation of C-band frequency combs with SiN deposited by plasma-enhanced chemical vapor deposition (PECVD), by replacing hydrogen- by deuterium-based precursors[36] to suppress absorption from N-H bonds. However, in this study, the comb is assumed to be generated on an external chip, which allows to filter out the pump and amplify the comb with a C+L band erbium doped fiber amplifier (EDFA) prior to coupling it into the main system chip. Complete integration of the two photonic chips is hindered in the configuration analyzed here by the low power conversion efficiency of bright soliton micro-ring cavities,[37] which leads to the requirement of an interposed optical amplifier. Higher conversion efficiencies, as provided by dark soliton generation[38] or advanced cavity design,[39] or the on-chip integration of optical amplification,[40] might, however, provide a path forward for a single-chip solution, which is the topic of a future investigation. The co-packaging of driver electronics with a flip-chip process is well established in silicon photonics[41] and can also be applied to SOH modulators provided the driver is shielded from the poling voltage during processing. This can for example be achieved with capacitive coupling.[42]

Figure 4 shows a system diagram of the PIC, with devices implemented in the SiN layer shown in green and devices in the silicon (Si) layer shown in blue. Due to the large optical power levels involved, the input edge coupler, the initial stages of the light splitting network, the reference branch in the Tx network, and the distribution network need to be implemented in SiN. The comb is assumed to carry 27 dBm in the fiber prior to be coupled in. However, active devices, such as SOH modulators in the Tx and PSs in the demodulator are implemented in the Si layer and see power levels in the few mW range, low enough not to experience significant nonlinear effects.[43]

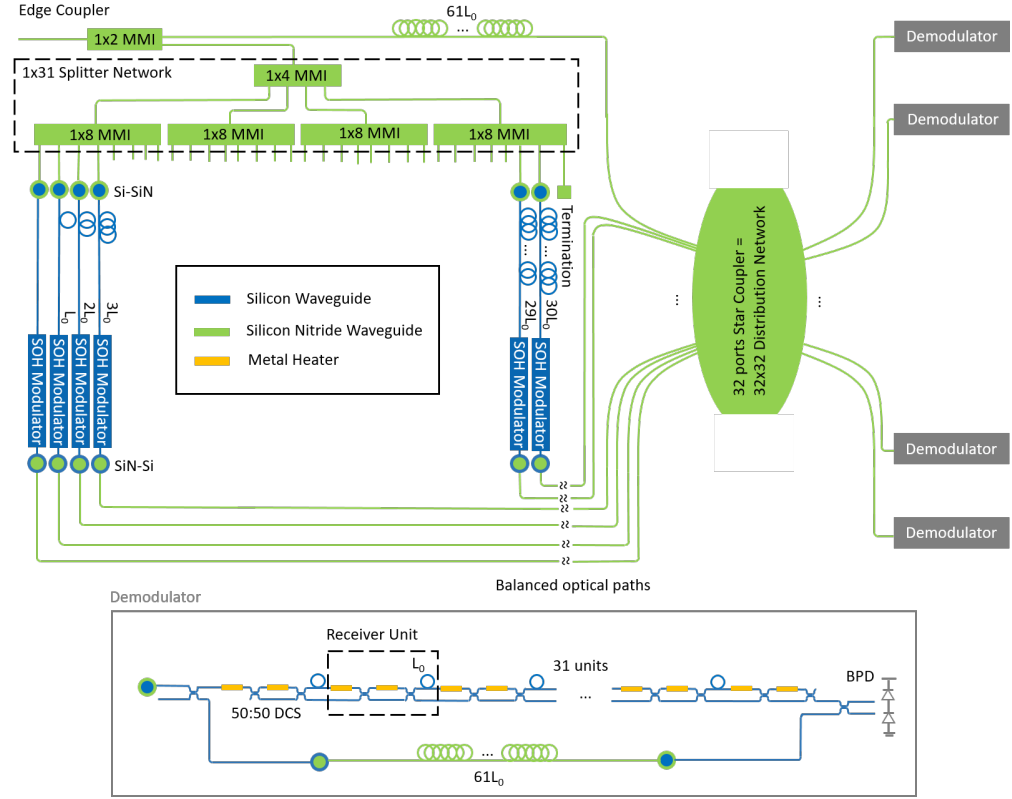

Fig. 4. Diagram of a 31×31 network. Devices defined in the SiN and Si layers are shown in green and blue, respectively. Delay loops required to shift signals and pilot tone between ODDM channels are shown explicitly.

Delay loops explicitly represented in the diagram allocate optical signals to specific ODDM channels. Balancing of other group delays and dispersion between the Tx and demodulator paths is critical to prevent signal leakage between logical channels. To minimize mismatch, matched delay lines implemented in the Tx and demodulator are implemented in the same layer, with nominally identical devices. Moreover, the additional group delay incurred in the Tx branches due to the splitter network and SOH modulators is compensated in the reference branch with a suitable compensation loop (not shown). Dispersion is balanced relative to the reference path by adding a short Si waveguide segment of reduced width (250 nm) carrying a less confined mode with opposite (normal) dispersion, with an equal length in each of the Tx signal paths. Similarly, group delays induced by the DCSs in the top branches of the demodulators are mirrored to their lower branches with suitable compensation loops, that are not shown in the diagram.

Not all device nonidealities can be corrected through the methods described above. Such are group delay mismatches between the star coupler ports, dispersion mismatches between the ports of the DCSs used in the demodulator, and the wavelength dependence of the SOH modulator efficiency, and are analyzed in Subsection 4.3.

The system occupies an estimated area below 100 mm$^2$, dominated by the demodulator layouts (55 mm$^2$). The largest single devices are the star coupler and 1x8 MMIs with footprints of ~400 μm by 200 μm and 50 μm by 640 μm, respectively. Assuming a comb FSR of 50 GHz and the minimum number of comb lines $Q = 4N - 1 = 123$ required for $N = 31$, the utilized optical spectrum spans $\Delta v = 6.15$ THz (49.3 nm). This results in a silicon delay line increment $L_0 = 11.1$ μm given a group index of 4.4 ($\tau_0 = 163$ fs). Since demodulation results from phase modulated ODDM carriers travelling through the lower branch of the demodulator, the signals are uniformly delayed by $61\tau_0 = 9.9$ ps between ANN layers, which is much below the clock cycle of state-of-the-art digital computing platforms, presenting an important advantage in latency even when compared to a single cycle. The area estimate has been done by extrapolating the size of a complete layout that has been sent to fabrication for a reduced scope 7-by-7 network. It includes all the photonic components, the space for the optical I/Os, and the pads required for flip-chip integrating the electronics for a single layer. Details of the floorplan estimate can be found in Appendix II.

In the following, the system impairments present in a practical implementation are individually analyzed with a numerical model. Methodological details of the physical model are reported in Appendix III.

### 4.1. Realistic comb shape and electronic transfer functions

Assumptions made in Appendix I were idealized to derive the orthogonality between ODDM channels mathematically. The electronics have been assumed to act as integrators perfectly gated over one unit interval (UI) and the comb to have a square shape. In this section, we assume a hyperbolic-secant-square shaped comb corresponding to micro-resonator-generated dissipative Kerr solitons[44] used as a light source without additional spectral shaping. Instead of an ideal integrator, TIA and modulator driver are modeled to have a continuous-time transfer function with a bandwidth limitation modeled by a fifth order Bessel filter. Ideally, a brick-wall filter with a cutoff frequency of half a comb FSR, i.e., the Nyquist frequency corresponding to the Baud rate given by the comb repetition time, would be used. The maximum system bandwidth of half an FSR would then be reached without compromising signal integrity, since the spectrum of each comb line would remain disjoint even after phase modulation. However, with a more realistic finite roll-off filter, a trade-off must be made between the cutoff frequency of the filter and residual overlap between the comb line spectra, resulting in inter-channel crosstalk.

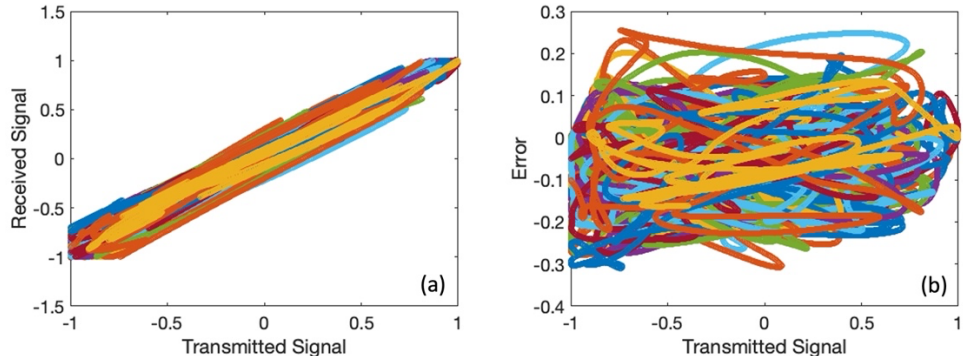

Fig. 5. Exemplary simulation result with a comb FWHM spanning 75% of $\Delta v$ and a Bessel filter cutoff frequency of 6.25 GHz. (a) Normalized demodulated photocurrent vs. transmitted signal for each of the 31 channels and (b) difference between the transmitted and the normalized received signals. Its std. dev. is $\sigma_\delta = 0.073$.

Figure 5 illustrates the methodology used to determine the signal integrity in the system. The simulation intentionally uses a secant-square shaped comb spectrum with a smaller full width at half maximum (FWHM) – 75% of the nominal $\Delta v$ – to induce significant signal distortion and inter-channel crosstalk. The x-axes of both panels show the transmitted signal amplitudes, defined here as $\sin(\varphi_n)$, $n \in [0, N-1]$, filtered by the same 5$^{th}$ order Bessel filter as the received photocurrent. Panel 5(a) displays the received differential photocurrents at downstream ANs, that are each tuned to a single logical channel. These photocurrents are divided by their maximum value to yield a dimensionless received signal varying between -1 and 1. While the signals can be seen to be transmitted and demodulated, the spread in the curves is indicative of signal distortion, which can be quantified by calculating the difference between the normalized received and transmitted signals, shown in panel 5(b). In this case, the std. dev. of this transmission error is $\sigma_\delta = 0.073$.

Figure 6 represents a systematic analysis of the impact of the comb FWHM and of the 5th order Bessel filter cutoff frequency on the signal quality. In panel (a), the Bessel filter cutoff frequency is fixed at 6.25 GHz, at which it is low enough to have a minimal effect on signal crosstalk, while the FWHM of the comb is varied. The results indicate that signal quality degrades rapidly for a FWHM below $\Delta\nu$. Even at FWHM = $\Delta\nu$, a significant penalty remains ($\sigma_\delta = 0.032$). It becomes negligible for a FWHM above $1.5\Delta\nu$ with $\sigma_\delta = 6.5\times10^{-3}$, which is much smaller than the penalty from shot and thermal noise derived below and does thus no longer limit the overall SNR. This corresponds to a FWHM of 75 nm, achievable using micro-resonator-generated dissipative Kerr solitons,[44] and is assumed in the following.

In panel (b), the FWHM is fixed at $1.5\Delta\nu$ and the cutoff frequency of the Bessel filter is varied. At the Nyquist frequency for 50 GBd operation, 25 GHz, the degradation is substantial due to the finite filter roll-off. However, at 12.5 GHz, the degradation is low ($\sigma_\delta = 0.018$) and becomes negligible below 7.5 GHz ($\sigma_\delta = 3.5\times10^{-3}$). The inset in panel (b) illustrates the transfer function of the 12.5 GHz cutoff Bessel filter. Above 25 GHz, it decays below -14 dB, so that a low level of spectral spill-over occurs between the modulated comb lines. At the same time, at the signaling rate of 25 GBd, signals below the 12.5 GHz Nyquist frequency are attenuated by less than 3 dB (electrical convention). This is the signaling rate considered for the overall system evaluation performed in Section 5.

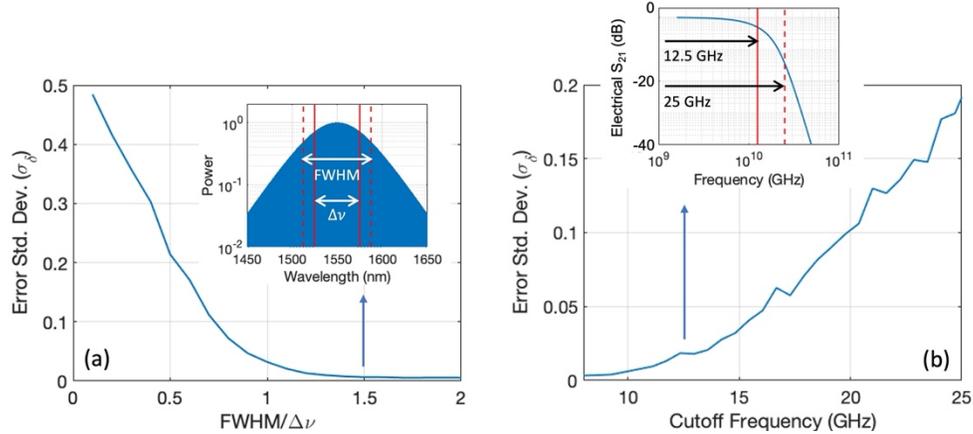

Fig. 6. Signal distortion (std. dev. of the difference between the demodulated photocurrent and the transmitted signal) versus (a) comb FWHM and (b) Bessel filter cutoff frequency. In (a), the Bessel filter cutoff is maintained at 6.25 GHz and in (b) the FWHM is maintained at $1.5\Delta\nu$ (75 nm). The inset of panel (a) displays the comb spectrum normalized to a peak comb line power of 1 and the inset of panel (b) shows the transfer function of the assumed Bessel filter.

### 4.2. Electronic and optical noise

Next, we analyze the effect of thermal noise, shot noise, relative intensity noise (RIN), optical linewidth and comb source jitter (radio-frequency (RF) linewidth).

The proposed ANN handles laser phase noise well due to the balanced homodyne detection scheme. The analysis distinguishes between two types of phase noise: correlated phase noise among all the comb lines, and uncorrelated phase noise associated with the RF linewidth of the comb and the jitter of the associated pulse train.[45,46] The actual demodulation of the signal is insensitive to both because the cumulative delay applied to the optical signal is exactly the same as that applied to the pilot tone, resulting in identical phase noise that cancels out during self-referenced coherent reception. However, the rejection of other signals, that are not matched to the delays applied in the demodulator, can be impacted because differential delays applied to ODDM carriers are then non-zero. They can be up to $\tau_\delta = (4N-2)\tau_0 = 19.9$ ps, the difference between the undelayed channel 0 traveling through the upper modulator branch and the pilot tone traveling through the lower modulator branch. Assuming a 1 MHz optical linewidth ($LW_{Opt}$, correlated phase noise) inherited from a pump laser in the comb generator, the std. dev. of the corresponding phase error is $\sqrt{2\pi LW_{Opt}\tau_\delta} = 0.64°$, which is very small. Numerical modeling confirms that this level of correlated phase noise does not have a significant impact on the network performance. However, a linewidth of 10 MHz or above would result in significant degradation, as shown in Fig. 7(a).

The comb's RF linewidth, $LW_{RF}$, converts into an equivalent jitter accumulated over the maximum differential delay $\tau_\delta$ as $\sqrt{2\pi LW_{RF}\tau_\delta}/2\pi\delta\nu$. For RF linwidths below 1 kHz, the resulting differential jitter is less than 1 fs, which is insignificant compared to $\tau_0$, 163 fs. Numerical modeling also confirms that RF linewidths up to a few kHz are tolerable (Fig. 7(a)). Phase noise measurements on RF signals from free running Kerr combs from SiN[47,48] or silica microtoroid[49] cavities indicate RF linewidths well below a kHz, indicating this not to be a limitation.

Differential signaling at the BPDs mitigates RIN, similar to other analog optical processing schemes.[11,50] A typical RIN level of -136 dB/Hz up to the 12.5 GHz electrical filter bandwidth results in a small penalty of $\sigma_\delta = 0.01$.

Next, we focus the analysis on the main noise limitations for this system architecture in the limit of high AN count, namely shot noise, that remains uncorrelated at the two BPDs, and thermal noise from electronics. We assume that a comb with 500 mW in the fiber is launched into the chip, and that various components of the system induce losses as follows: 2 dB for each of the edge coupler and the splitter network at the beginning of the Tx subsystem, 3 dB for the demodulator, and 2 dB for the overall waveguide routing. The star coupler is assumed to have 5 dB excess losses, based on the worst performing channels of the device described in Appendix II, in addition to the nominal splitting losses. We model phase modulation using high-speed SOH phase shifters,[51] sized to result in a $\pi$ phase shift with a 2 V$_{pp}$ signaling scheme, resulting in 2.55 dB insertion losses. The BPD are assumed to have a responsivity of 0.8 A/W and the TIA to have a low input referred noise current density of $I_n = 10\ pA/\sqrt{Hz}$ enabled by co-design with low capacitance waveguide photodiodes.[52]

Shot noise is also a dominant limitation in this architecture due to the high average power received by the BPD, which is different from conventional short-distance communication links. In this case, even when only one logical channel is demodulated, the power of all other channels arrives at the BPD and is converted into a common mode photocurrent. Since shot noise at the two BPDs is uncorrelated, it has to be evaluated based on the total generated photocurrent.

Assuming a FWHM of $1.5\Delta\nu$ and a 12.5 GHz Bessel filter 3-dB cutoff, a demodulated differential current of ±41.9 µA is simulated when the demodulator is tuned to one channel only. The simulated shot ($\sigma_{Sh}$) and thermal noise ($\sigma_{Th}$) std. dev. are well in line with analytical expressions given by

$$\sigma_{Sh} = \sqrt{2qI_{cm}f_{NEB}} = 1.47\ \mu A \tag{3}$$

$$\sigma_{Th} = I_n\sqrt{f_{NEB}} = 1.13\ \mu A \tag{4}$$

where $q$ is the elementary charge, $I_{cm}$ is the average common-mode photocurrent generated at the BPDs (532.7 µW), and $f_{NEB}$ is the noise equivalent bandwidth (NEB) of the Bessel filter. Together with the distortion induced by the finite Bessel filter bandwidth, these result in an overall $\sigma_\delta$ of 0.05 (cf. Fig. 7(b)). While this evaluation was done with the demodulator tuned to one channel only, it is also characteristic of the overall signal quality in the general case after summation at the receiving AN of uncorrelated channels, since the overall received signal strength is then in the same order, see Section 3.

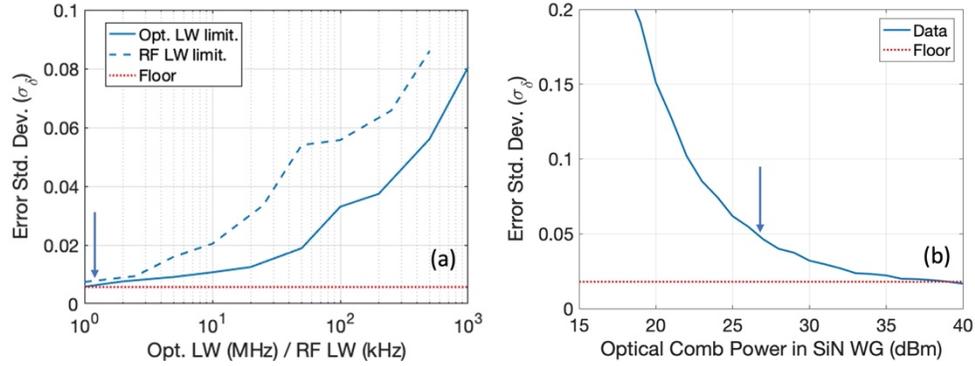

Fig. 7. Simulated noise std. dev. $\sigma_\delta$ versus (a) assumed optical and RF comb source linewidths and (b) optical comb source power. In both (a) and (b), the comb FWHM is set to 1.5Δν. In (a), the Bessel filter cutoff is set to 6.25 GHz and shot noise and thermal noise are ignored to maintain a low noise floor, allowing to estimate the impact of the linewidths on their own. In (b), these noise sources are turned on. Moreover, the target scenario with a 12.5 GHz Bessel cutoff frequency is assumed to obtain the corresponding NEB. This includes the higher distortion floor established in Section 4.1. Arrows in (a) and (b) indicate the operating conditions assumed in the following.

### 4.3. Group velocity mismatch, dispersion, and spectral modulation efficiency

The proposed network requires optical devices with wideband operation. In the ideal network, group delays are balanced, except for those introduced explicitly for signal translation between ODDM carriers. Additionally, the same amount of cumulative dispersion is applied to all optical paths, and the phase modulation applied by the high-speed phase modulators is constant across wavelength. Actual device designs deviate from these assumptions, and we analyze such deviations here.

We assessed the level of tolerable group delay mismatch by adding group delay offsets (Δτ) to individual Tx-subsystem signal paths, and the results are presented in Fig. 8(a). Signal degradation remains modest up to $\Delta \tau = 50$ fs, degrades rapidly around $\Delta \tau = 80$ fs, and signals become fully scrambled for $\Delta \tau > 100$ fs. This result was expected from the 163 fs $\tau_0$, as at $\Delta \tau = 80$ fs, roughly half the signal power from a given logical channel is shifted to the adjacent one, leading to severe crosstalk. At $\Delta \tau = 160$ fs, an entirely different, uncorrelated channel is demodulated. Physically, group delay mismatches result from various sources.

As described above, dispersion can be largely balanced out. An exception is the dispersion induced by the incremented silicon delay lines introduced in the Tx signal branches and the upper demodulator branch, since these are of varying length. Ideally, these delay lines should be free of dispersion, which can be obtained with a 380 nm wide and 220 nm thick Si (fully etched) ridge waveguide. This is slightly narrower than the common single mode interconnect waveguide width used in 220 nm SOI technology. A 400 nm wide waveguide results in a dispersion parameter of *D* = 660 ps/nm/km, which is small enough not to be a significant issue because crosstalk primarily arises from adjacent ODDM channels, which is determined by the dispersion accumulated along a single length increment $L_0$. The maximum group delay mismatch of ±0.3 fs accumulated across the 75 nm FWHM of the comb is confirmed by numerical modeling to have an unsignificant impact on network performance.

Another factor that can affect the network is the wavelength dependency of the phase modulator efficiency, which is caused by the dispersion of the nonlinear r33 coefficient[53] of the utilized nonlinear polymer and the wavelength dependence of the field overlap with the waveguide slot.[54] Fortunately, these have opposite signs in the chosen design and partially cancel out. A 4.2% efficiency change is estimated across the 75 nm FWHM of the comb, resulting in a data dependent group delay of 2.3 fs when a full $\pi$ phase shift is applied. This is much smaller than $\tau_0$ and thus also results in a negligible penalty.

The group delay mismatches introduced by the initial splitter network in the Tx, relying on 1x4 and 1x8 MMIs, are also small (±4 fs across all ports) and can be compensated, since each Tx branch processes a single signal. However, the star coupler introduces larger mismatches, up to ±20 fs for light injected through off-center ports (Appendix II). Since these depend on both the input and output port index, they cannot be easily corrected. At lower port counts, the distribution network could be implemented instead by an MMI, for which techniques have been devised to obtain broadband operation[55] with low phase errors.[56] However, for the high port counts considered here, a star coupler appears more practical.

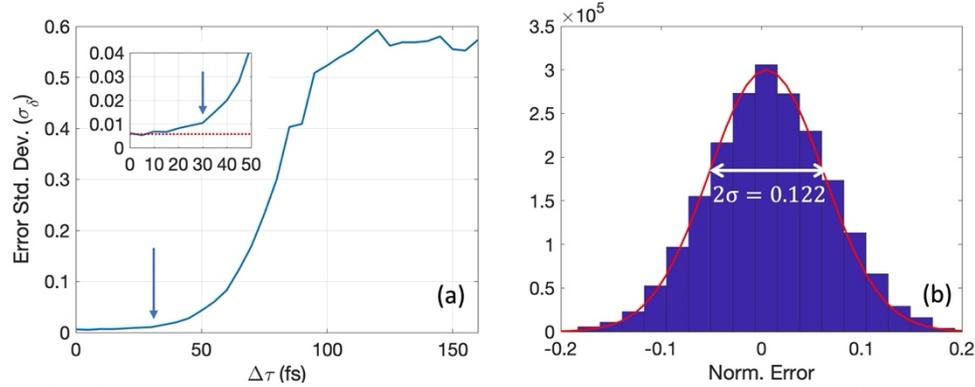

Fig. 8. (a) Simulated noise std. dev. $\sigma_\delta$ versus group delay mismatch applied in the Tx branches. The Bessel filter cutoff is set to 6.25 GHz and the noise sources turned off, to obtain a low noise floor allowing to estimate the impact of the group delay mismatch alone. (b) Histogram of the demodulated signal error with the Bessel filter cutoff set to the nominal 12.5 GHz, noise sources turned on, and device nonidealities considered.

A similar challenge results from the DCSs used in the upper demodulator branch, that have been implemented as wide bandwidth devices following the concept of Lu *et al*.[57] For such 2×2 devices, the group delays can be balanced out for all four optical paths at the comb's center wavelength, since there is an equal number of ports to which waveguide segments can be added for compensation. However, the dispersion is different for the $S_{31}$ (up-up) and $S_{42}$ (down-down) terms of the transfer matrix, see Appendix II. While a single DCS applies a very small penalty, 62 of them are cascaded in each demodulator leading to substantial mismatches. The cumulative penalty arising from the DCSs, star coupler, SOH modulator, and waveguide dispersion results in $\sigma_\delta = 0.012$, corresponding in an effective group delay mismatch of ±30 fs (cf. Fig. 8(a)). With the electronic filter bandwidth increased back to 12.5 GHz and shot and thermal noise turned on, the overall signal quality degrades to $\sigma_\delta = 0.061$ (all simulated jointly).

This corresponds to a signal quality (Q-)factor of 16.4, so that binary data could be easily transmitted error free. However, for the analog signaling scheme considered here, the relevant metric is the SNR, which is evaluated as 21.3 dB and corresponds to an effective number of bits (ENOB) of 3.2 for a single AN-AN link. This is the link-SNR assumed in the next section for a fully populated network (31×31), where noise is added as random Gaussian noise, as physical level modeling would be prohibitively computationally expensive for iterative network training. Fig. 8(b) shows a histogram of the deviation between transmitted and decoded data, with the overlaid Gaussian noise model. Further scaling of the network is limited by the factors analyzed above. These are mainly (i) the spectral width of available comb sources and the availability of devices operating over a sufficiently wide wavelength range, (ii) the available optical power limited by the damage threshold of the SiN waveguide at the beginning of the splitter network, and (iii) cumulative phase errors as an increasing number of devices is cascaded inside the demodulators with increasing channel count. Since these factors have been co-optimized in the network evaluated here, each of them contributes with a non-negligible amount to the overall signal degradation, however, the main limitation remains the thermal and shot noise limited SNR caused by the finite optical power.

A more quantitative look can also be taken at the expected temperature sensitivity of the network, since it results from the imbalances of the optical path lengths evaluated above. Taking ±30 fs for $\Delta\tau$, as evaluated above, and assuming that ±15 fs arise from the silicon part of the network, which has the larger thermo-optic coefficient, we estimate the temperature dependent phase offset applied to the phase encoded data from

$$\frac{\partial \varphi}{\partial T} = \omega \Delta \tau \frac{1}{n_g} \frac{\partial n_{eff}}{\partial T} \qquad (5)$$

From this, we estimate a phase error of less than 0.08 radian over a 100 ºC temperature swing. To put this in context, this results by itself in $\sigma_\delta = 0.057$ and is of comparable magnitude as the noise penalties. In an experimental context, the temperature sensitivity will of course also depend on how repeatably devices are fabricated across the network.

Another temperature sensitivity in the system arises from the mismatch in thermo-optic coefficients between the SOH modulators and silicon compensation loops in the Tx signal branches on the one hand and the corresponding SiN compensation loops in the Tx reference branch om the other hand. To address these, a single PS can be added to the Tx reference branch, enabling nearly athermal operation for the rest of the system, for which the use of different materials is balanced between the optical paths. Alternatively, these compensation loops can also be duplicated in the respective branches of the demodulators, which will then rebalance the thermal dependance by mirroring it in the other optical path and thus cancel it.

## 5.  Nonlinear equalization of an optically enabled time-interleaved ADC

To test the effectiveness of the proposed ANN in practical applications, we evaluate its performance as a nonlinear equalizer in the presence of signal distortion and noise, using the overall system ENOB as a benchmark. Digital electronic ANNs have already been successfully applied to such nonlinear equalization tasks in both short-reach fiber links[58] and long-haul communications.[59] In this study, we apply nonlinear equalization to the output of the optically enabled (OE), time-interleaved (TI-)ADC architecture shown in Fig. 9(b), for which we have previously benchmarked linear feed-forward equalization (FFE) with digital electronics.[60]

The OE-TI-ADC samples an electrical signal by applying it to a MZM, to which pulses with different center wavelengths are being fed.[11] Their separation by center wavelength at a following optical processing stage subdivides the modulated pulse train into a number of reduced rate sample streams, that can be analyzed by lower speed electrical ADCs. However, time- and frequency-domain signal leakage between the pulse trains results in signal distortion, which was previously addressed with FFE.[60] The MZM must also be driven in the small signal regime to prevent nonlinear distortion of the sampled data, limiting the SNR of the system.[61] To address this limitation, the FFE is replaced by an ANN that allows nonlinear signal equalization and thus the driving of the MZM with a higher signal strength, reducing the number of optical amplifiers required to maintain the SNR. Signal processing in the analog domain, by interposing the ANN between the photoreceivers of the OE-TI-ADC and the digital electronics, allows performing the nonlinear equalization before quantization noise is incurred, so that amplification of quantization noise by the equalizer can be avoided.[60] The overhead associated with implementing the equalization with an OEO-ANN is reduced since integrated photonics are already required for the front-end of the OE-TI-ADC, and the same comb source can be used for both systems. While a higher pulse repetition rate is required for the ANN to avoid aliasing, similarly to what happens in asynchronously clocked time-interleaved electronic architectures,[62] the doubling from 25 GHz to 50 GHz FSR can be straightforwardly obtained by using an interleaver (imbalanced MZI) selecting every second comb line prior to optical amplification. This interleaver can be integrated together with the SiN ring resonator on the auxiliary chip generating the comb (Fig. 9(a)).

The system schematic in Fig. 9 shows the implementation of six interleaved channels in the OE-TI-ADC front-end, with an overall sampling rate of 150 GS/s and a conversion bandwidth of 75 GHz. Each channel is equalized by an independent ANN under consideration of the data from the other channels. The MZM is driven with a high signal strength reaching 80% of the maximum range (corresponding to $\pm\pi/4$ phase shift in push-pull operation), with the resulting nonlinearity compensated by the ANN. Consequently, two booster optical amplifiers (BOA), that would be otherwise necessary to amplify the complementary outputs of the MZM,[60] have been removed from the network model.

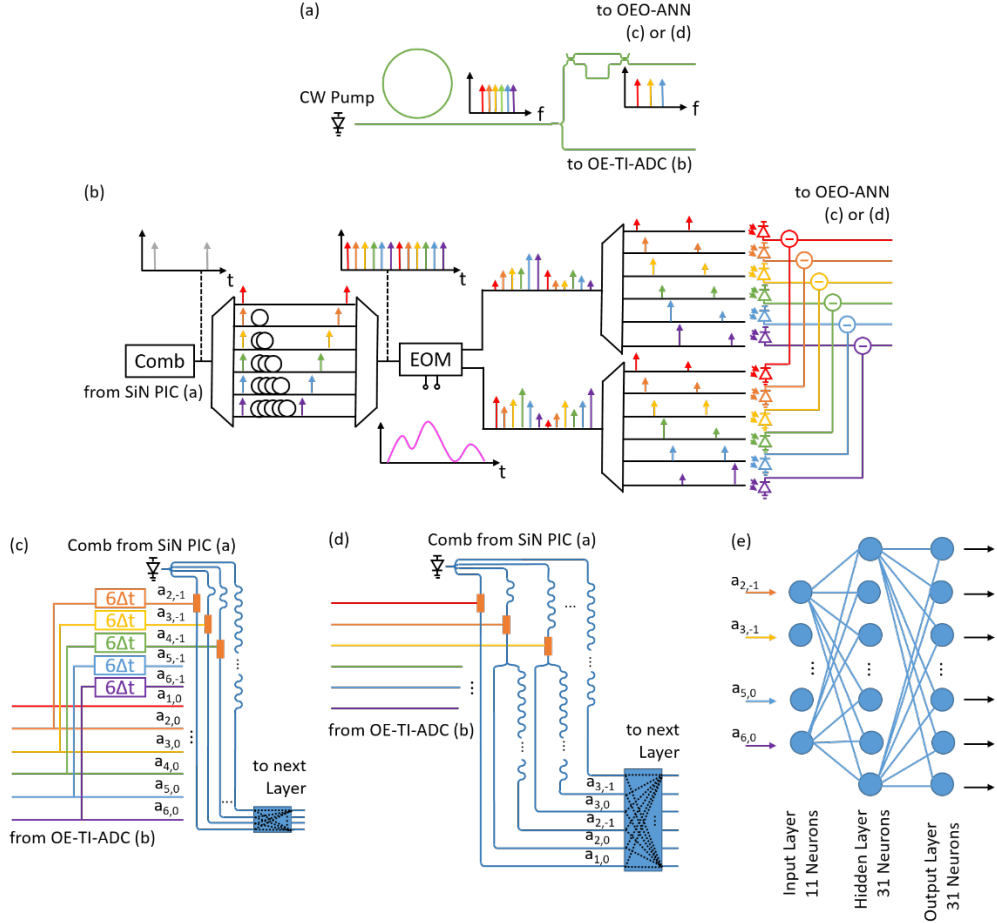

Fig. 9. (a) Schematic of the auxiliary SiN chip generating the 25 GHz comb shared between the OE-TI-ADC front-end and the OEO-ANN. Prior to routing it to the ANN, the FSR of the comb is doubled with an imbalanced MZI. (b) 6-channel OE-TI-ADC front-end. Interleaved pulse trains with different center frequencies are fed to a dual-output MZM, after which the pulse trains are de-interleaved by a wavelength division demultiplexer. (c), (d) Schematics of the ANN input layer. In (c), required delays are implemented in the electrical domain, prior to electro-optic phase modulation. In (d), signals are split and delayed in the optical domain instead, with delays chosen such that the delayed optical signals are mapped to free ODDM channels. (e) Higher-level schematic of the overall ANN, with two optical networks connecting respectively the input to the hidden layer (11×31) and the hidden layer to the output layer (31×31), whose output can be directly decoded with binary electronics.

For each OE-TI-ADC channel, an 11-tap nonlinear equalizer is implemented by feeding the analog waveform of the five pulses preceding and following the equalized pulse to the ANN, wherein the 40 ps time delays required to process all these samples in the same ANN clock cycle are assumed to be implemented in the electrical domain (see Fig. 9(c), in which the equalized channel is color-coded in red). This ANN then recovers one of the subsampled versions of the input signal with a 25 GS/s sampling rate, that is subsequently combined with the output of the other ANNs to generate the 150 GS/s sample stream.

Alternatively, such a delay can be implemented in the optical domain, by splitting the light after the phase modulators of the input layer and delaying one branch with a ~2.7 mm Si delay line (Fig. 9(d)). Since 40 ps correspond to a multiple of the inverse comb FSR in the ANN network, this delay reallocates an ODDM channel to itself, leading to two waveguides carrying data in the same ODDM channel being combined at the distribution network, and to unwanted interference to occur. Instead, the time delay can be slightly modified to 40 ps + $\tau_0$ to allocate the delayed samples to an adjacent ODDM channel purposefully left free, allowing parallel processing. This does, however, further burden the optical power budget and the complexity of the optical system. The simpler architecture shown in Fig. 9(c) has instead been numerically analyzed in the following.

The hidden ANN layer processes the 11 input-streams with 31 ANs that feed the data to a second 31-neuron output layer (Fig. 9(e)). The ANN is trained such that an output AN of index $p \in [0,30]$ outputs a Boolean value that switches to one if the input waveform exceeds a certain threshold. This threshold corresponds to $(p + 0.5)LSB$, with $LSB$ the least significant bit of the OE-TI-ADC, providing the equivalent function of a five-bit flash ADC. The OE-TI-ADC output can then be obtained by simple digital summation.

The hidden and the output layer both use a sigmoid activation function defined to have rails at 0 and $V_\pi$, with $V_\pi$ the voltage required to induce a $\pi$ phase shift in the following phase modulator. The overall transfer function of the TIA + driver + phase modulator is defined such that a zero differential input photocurrent is mapped to $V_\pi/2$, around which the amplification chain has a small signal gain of 10 rad/mA. The input layer of the network is also assumed to receive signals scaled such that the main equalizer tap (AN #5 color coded in red in Fig. 9(c)) receives a signal in the range $[0 \ V_\pi]$. The simplified demodulator architecture shown in Fig. 2(d) is assumed. While it restricts the programmable parameter space to some extent, it also halves the number of PSs required.

A second network architecture with reduced functionality is studied in parallel, where the output layer is replaced by a single output neuron implementing an identity activation function. The purpose of this architecture is to serve solely as nonlinear equalizer without digitization. In this case, the target is to generate an output that accurately reconstructs the subsampled input signal of the OE-TI-ADC while compensating for signal leakage between the OE-TI-ADC channels and for the nonlinear transfer function of the MZM sampler.

To train the networks, we use 57 waveforms at frequencies ranging from 5 to 75 GHz, with each waveform consisting of 97 samples, for a total of 5529 training samples. We employ a stochastic gradient descent training method reiterated over 60 EPOCHs. A backpropagation algorithm with an ADAM optimizer directly trains the PSs determining the ANN weights, since the latter are not accessible directly. As the two investigated networks have different objectives, they are trained independently with different loss functions. They are further referred to as the analog and digital output ANN. To have a differentiable output based on which to evaluate the loss function in the case of the digital output ANN, the latter is applied to its analog outputs prior to the digital thresholding as a mean square error summed over all the output neurons. However, to improve the quality of the training result, this output is preconditioned by a sigmoid of increasing steepness as part of the loss function evaluation.

Figure 10(a) presents the output photocurrent of the analog output ANN as a function of both the input to the ANN (input AN #5 transducing the main tap of the equalizer) and of the input to the OE-TI-ADC. The plot shows that the output versus OE-TI-ADC input is linear with small scatter, indicating high-quality signal reconstruction. However, when plotting the output versus the ANN input, the data appears as a point cloud that exhibits both scatter and an overall nonlinear transformation, as revealed by the shape of the polynomial fit. These characteristics correspond to the desired functionalities since the nonlinear distortion compensates for that of the MZM (as evidenced by the curve's curvature, opposite to that of an MZM). Moreover, the broadening of the point cloud reveals the necessity of correcting the data by taking into account the other taps, which is adequately done by the network as seen in the overall OE-TI-ADC+ANN transfer function.

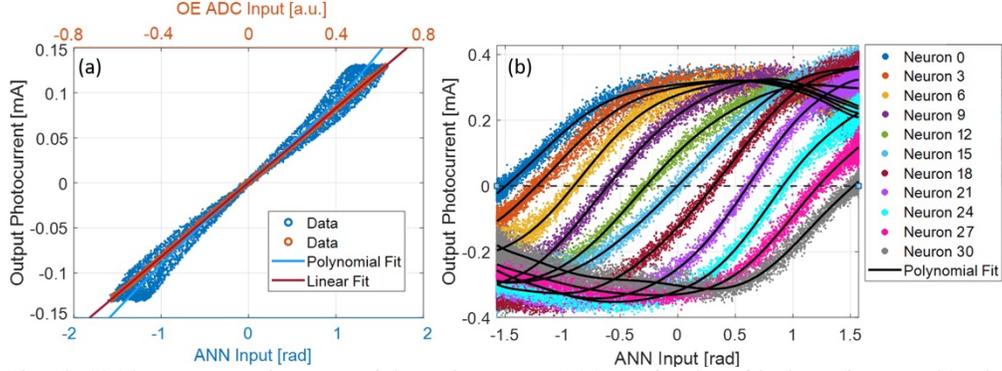

Fig. 10. (a) Photocurrent at the output of the analog output ANN as a function of its input (bottom axis). The nonlinearity of the system is apparent. The ANN corrects for the OE-TI-ADC front-end nonlinearity, as shown by the red curve. It corresponds to the output of the ANN as a function of the input to the OE-TI-ADC (top axis) and has a very linear profile. In (b), the photocurrent of every third output neuron of the digital output ANN is shown as a function of the input to the ANN. The crossing of these curves with the zero differential photocurrent level corresponds to the switching threshold of the following 1-bit digitizer.

In Fig. 10(b), the output photocurrent of the digital output ANN is shown for every third output AN, prior to applying the last sigmoid and the thresholding operation. The dashed line represents the decision threshold of the subsequent single bit digitization stages. The AN-responses cross that threshold at different levels of ANN input, determining the switching point of the digital output. Unlike the analog output ANN, in which the entire analog response is constrained, here only this threshold matters. Therefore, the ANN can maximize the gain of the signal processing chain through increasing of the weights, even if this leads to a strongly nonlinear demodulator response. The sine-curve shaped response of the output ANs is visible in Fig. 10(b), and the differential output currents vary in the maximally achievable output range close to ±0.4 mA. In contrast, the response of the first ANN requires a small level of well-controlled nonlinearity and is constrained to a smaller signal regime achieved through weight reduction during training. As a consequence, its output response is about a factor 3 weaker. This has important consequences for the tolerance to noise, as we shall see in the following.

The performance of the network is evaluated both with and without Gaussian additive noise during training and performance validation. The noise associated to the 31×31 network between the hidden and output layers has been determined in the previous section. However, the 11×31 network between the input and hidden layer has a higher SNR due to the reduced number of upstream ANs, resulting in a $\sqrt{31/11} \simeq 1.7$ times larger signal strength at equal optical power per network. Thermal and shot noise levels remain unchanged, for the latter since an equal amount of optical power is distributed over an equal number of downstream ANs (31).

Figure 11(a) shows the network validation results, where the SNR and ENOB of the ANN outputs are evaluated at every signal frequency. Results are shown for both network types, with and without noise for the digital output ANN, and compared to the unequalized output of a single OE-TI-ADC channel. The stark drop in performance of the unequalized channel is due to increased inter-channel signal leakage at higher signal frequencies.[60] At the same time, the performance is also bounded by the nonlinear distortion introduced by the MZM sampler, which determines the SNR at lower signal frequencies. In the absence of ANN network noise, the single output ANN equalizes the data very well, achieving an ENOB of 5.5. However, when the network noise is considered, the performance drops significantly, to an ENOB of 3.6. The sensitivity to noise of this network is further investigated in Fig. 11(b), in which noise levels are rescaled relative to the nominal case. The ENOB averaged over all signal frequencies is plotted as a function of the rescaling factor. It is apparent that the SNR of this network drops 6 dB for every doubling of the noise std. dev., in line with noise-limited performance.

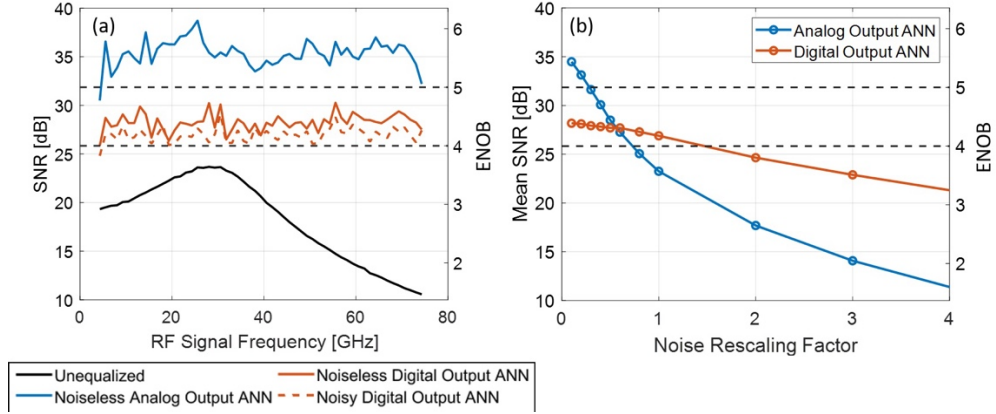

Fig. 11. (a) SNR levels of the processed signals in different scenarios, highlighting the improvement with respect to the non-equalized signal, also in presence of noise. The SNR is assessed by processing sinusoidal signals at the frequencies reported on the abscissa. In (b) the mean SNR, assessed by averaging the variances of signals at multiple frequencies, is shown as a function of different rescaling factors applied to the std. dev. of the noise terms. The analog output ANN is seen to be significantly more sensitive to noise than the digital output ANN.

The situation is markedly different for the digital output ANN. Since this network does the digitization itself and reports numbers between 0 and 31, the maximum achievable ENOB, as limited by the actual number of bits, is 5. The actual performance is slightly above 4 bits. This drop in performance is attributed to the restricted programmability of the simplified demodulators, as verified for example by the near perfect performance of a 4-layer network. In particular, the simplified demodulators restrict the ability of the network to self-bias itself, as enabled by the full demodulator shown in Fig. 2(c). However, the tolerance to noise of the digital output network is much better, as a consequence of the higher signal gains in the network and the binary thresholding operations made at the output. The performance with the nominal noise level is almost as good as noiseless performance, see Fig. 11(a). Rescaling this noise to higher levels in Fig. 11(b), it is also visible that the resulting performance curve is relatively flat, indicating that the network is limited by both noise and training / the available degrees of freedom.

It should be noted that in general the ENOB of the whole system is not limited by that of a single link: Even in the case of a single analog output, the system-level SNR can exceed that determined for a single link in Section 4, since several hidden layer ANs can collectively contribute to the propagation of the same signal. In the case of the digital output ANN, the system-level performance is improved due to the high gain of the network transfer function around the decision threshold, enabled by the single bit decision made at every output. The link-level ENOB can however serve to derive the equivalent number of operations performed by the digital output ANN: $11 \times 31 + 31 \times 31 = 1302$ three-bit additions every 40 ps, corresponding to over $32 \times 10^{12}$ operations per second. A quantitative benchmarking against an all-digital implementation will be the object of a future study.

## 6. Conclusions

Our study presents a novel scalable optical-electrical-optical artificial neural network concept, which has been evaluated under conditions of noise and distortion. The modeling results demonstrate that the proposed architecture can support a $31 \times 31$ network, which equates to a remarkable 961 point-to-point on-chip interconnects and programmable weights, while utilizing a single 500 mW comb. The signal provides a signal-to-noise ratio of 21.3 dB and an effective number of bits of 3.2 for each neuron.

To evaluate the applicability of our concept to practical signal processing tasks, we modeled the nonlinear equalization of a signal generated by an optically enabled time-interleaved ADC architecture, confirming the trainability of our network and obtaining a system-level effective number of bits over 4 over the entire 75 GHz ADC bandwidth. We achieved these results by introducing a novel orthogonal delay-division multiplexed signaling scheme with pilot tone based self-homodyne detection, which improves the scaling of the optical power budget. Once trained, it also enables stable uncooled operation without retraining in a dynamic thermal environment.

Our team designed a set of optically wideband integrated devices specifically for this purpose into an openly accessible 220 nm silicon-on-insulator silicon photonics technology that supports back-end-of-line silicon nitride waveguides with a higher power handling capability.

We anticipate that our network architecture will enable low latency and ultra-broadband nonlinear signal processing in high performance optically enabled data converters and real-time control systems.

**Appendix I: Orthogonality conditions and spectral efficiency**

We assume a square shaped comb with $Q$ comb lines. During ODDM channel allocation, a delay $QL_0$, with $L_0 = c_0/\Delta\nu n_g$, is equivalent to a zero delay, due to the periodic nature of the comb emission. Consequently, we require $Q \geq 2N$ for the architecture shown in Fig. 1(a). The longest delay applied by an upstream AN is $(N-1)L_0$, resulting in a cumulative delay $(2N-1)L_0$ after travelling through the lower demodulator branch that applies a delay $L_R = NL_0$ in that case. This requirement becomes $Q \geq 4N - 1$ in the architecture shown in Fig. 1(b), since the longest cumulative delay applied to the pilot tone when travelling through the lower demodulator branch is then $2L_R = (4N-2)L_0$.

As explained below, the electronic part of the ANs needs to suitably lowpass filter or integrate the differential photocurrent for the beat tones resulting from comb lines of different frequency not to be recorded. The smallest gating time that satisfies this requirement, which is also the UI of the signals, is the repetition time of the comb source, $T_{rep} = 1/\delta\nu$ (but it can also be a multiple thereof). This corresponds to a modulation Nyquist frequency of $\delta\nu/2$. Given that the assumed phase modulation results in dual sidebands, with Nyquist modulation this would fill the available optical spectrum, but without overlapping of spectra generated from individual comb lines. The aggregate system bandwidth, corresponding to the signal bandwidth per channel times the total number of channels, thus corresponds to 1/4th of the comb's spectral width in Fig. 1(a), due to dual sideband modulation and the number of available logical channels being in the order of $Q/2$, i.e., half the ODDM channels remain unused. In case of Fig. 1(b), the overall system bandwidth drops to 1/8th of the comb spectral width, resulting from only a quarter of the ODDM channels being used for data transport.

The initial spectrum of the dispersed comb, with an FSR $\delta\nu$ and an overall spectral extent $\Delta\nu$, is given by

$$E_c = \sum_{q=0}^{Q-1} E_0 e^{i(\omega_q t + \theta_q)} \tag{6}$$

with $Q = \Delta\nu/\delta\nu$ the number of comb lines, $E_0$ their amplitude, $\omega_q = \omega_0 + 2\pi q \delta\nu$ their angular frequency, and $\theta_q$ their phase. After phase modulation by the different ANs and superposition in the distribution network, the field arriving at the demodulator of the receiving neuron of index $p$ is

$$\frac{10^{-\frac{IL_R}{20}}}{\sqrt{2N}} \sum_{q=0}^{Q-1} E_0 e^{i(\omega_q t + \theta_q + \gamma_{pR} - \omega_q \tau_R)}$$

$$+ \frac{10^{-\frac{IL_R + IL_{mod}}{20}}}{\sqrt{2N}} \sum_{n=0}^{N-1} \sum_{q=0}^{Q-1} E_0 e^{i(\omega_q t + \theta_q + \gamma_{pn} + \varphi_n(t) - n\omega_q \tau_0)} \quad (7)$$

where $\gamma_{pn}$ and $\gamma_{pR}$ are phases introduced by the distribution network (that are unavoidable to obtain a unitary transfer function[32]). $\tau_0 = n_g L_0/c_0$ and $\tau_R = n_g L_R/c_0$ are the group delays associated to $L_0$ and $L_R$. To reduce mathematical expressions to manageable sizes, we are restricting the following derivation to considering only the modulated optical signals in the lower branch of the demodulator and the pilot tone in its upper branch. The derivation for the other terms can be straightforwardly obtained following the same steps so as to show, there too, the required orthogonality. The complex valued amplitudes of the superposed optical signals at the end of the lower branch, $E_p$, and of the pilot tone at the end of the upper branch, $E_R$, are given by

$$E_p = \frac{10^{-\frac{IL_R + IL_{mod}}{20}}}{2N} \sum_{n=0}^{N-1} \sum_{q=0}^{Q-1} E_0 e^{i(\omega_q t + \theta_q + \gamma_{pn} + \varphi_n(t-\tau_R) - \omega_q(n\tau_0 + \tau_R))} \quad (8)$$

$$E_R = \frac{10^{-\frac{IL_R}{20}}}{2\sqrt{N}} \sum_{q=0}^{Q-1} E_0 e^{i(\omega_q t + \theta_q + \gamma_{pR} + \eta_{pm} - \omega_q(m\tau_0 + \tau_R))} \quad (9)$$

where $m$ is the index of the received logical channel and corresponds to the delay $mL_0$ in the upper interferometer branch (the general case can be expressed as a summation over $m$). $\eta_{pm}$ is the additional phase applied to the pilot tone in the upper demodulator branch in the optical path associated to this delay. This results in the differential photocurrent

$$I_p = 2R \cdot \mathrm{Re}(-iE_p E_R^*) = \frac{10^{-\left(\frac{IL_R}{10} + \frac{IL_{mod}}{20}\right)} R}{2N\sqrt{N}} \mathrm{Re}$$

$$\left[-i \sum_{n=0}^{N-1} \sum_{q=0}^{Q-1} \sum_{q'=0}^{Q-1} |E_0|^2 e^{i\left((q-q')\delta\omega t + (\theta_q - \theta_{q'}) + (\gamma_{pn} - \gamma_{pR} - \eta_{pm}) + \varphi_n(t-\tau_R) - (n\omega_q - m\omega_{q'})\tau_0\right)}\right] \quad (10)$$

This equation can be easily simplified if the differential photocurrent is integrated over a UI given by $1/\delta\nu$ or a multiple thereof, as previously assumed, and if the modulated phases $\varphi_n$ are kept constant over the entire gating period. In that case, we use the orthogonality condition

$$\frac{1}{T_{UI}} \int_0^{T_{UI}} e^{i(q-q')\delta\omega t} dt = \delta_{q,q'} \quad (11)$$

so that Eq. (10) reduces to

$$I_p = \frac{10^{-\left(\frac{IL_R}{10} + \frac{IL_{mod}}{20}\right)} RP_c}{2N\sqrt{N}Q} \mathrm{Re}\left[-i \sum_{n=0}^{N-1} \sum_{q=0}^{Q-1} e^{i\left((\gamma_{pn} - \gamma_{pR} - \eta_{pm}) + \varphi_n(t-\tau_R) - (\omega_0 + q\delta\omega)(n-m)\tau_0\right)}\right]$$

$$= \frac{10^{-\left(\frac{IL_R}{10} + \frac{IL_{mod}}{20}\right)} RP_c}{2N\sqrt{N}Q} \mathrm{Re}\left[-i \sum_{n=0}^{N-1} \left(e^{i\left((\gamma_{pn} - \gamma_{pR} - \eta_{pm}) + \varphi_n(t-\tau_R) - i\omega_0(n-m)\tau_0\right)} \sum_{q=0}^{Q-1} e^{-iq\delta\omega(n-m)\tau_0}\right)\right] \quad (12)$$

Since the condition imposed on $L_0$ to be a multiple of $c_0/n_g\Delta\nu$ results in $\delta\omega\tau_0$ to be an integer multiple of $2\pi/Q$, the summation over $q$ can be recognized as being a discrete Fourier transform (DFT) of a string of ones, resulting again in a Kronecker delta $Q\delta_{m,n}$ and

$$I_p = \frac{10^{-\left(\frac{IL_R}{10}+\frac{IL_{mod}}{20}\right)}RP_c}{2N\sqrt{N}} \sin\left(\left(\gamma_{pm} - \gamma_{pR} - \eta_{pm}\right) + \varphi_m(t - \tau_R)\right) \tag{13}$$

which corresponds to Eq. (1), with the phase offsets introduced by the distribution network for both the signal and pilot tone and the group delay introduced by the delay loop in the demodulator taken into account.

This communication scheme can be seen to have analogies with optical orthogonal frequency division multiplexing (OFDM)[63] and shares some of the underlying formalism, but instead of encoding data by applying it to several time domain samples varying in time according to a subchannel frequency, the data is applied to several comb lines whose complex valued amplitudes vary with frequency according to an ODDM channel time delay.

Optical signals and their superpositions can either be represented as a vector of comb line amplitudes indexed by the comb line index $q$, $\boldsymbol{v} = [v_q]$, or by a vector of logical channel amplitudes indexed by the logical channel index $n$, $\widetilde{\boldsymbol{v}} = [\widetilde{v}_n]$, wherein $\boldsymbol{v}$ is transformed into $\widetilde{\boldsymbol{v}}$ (and vice versa) by operating an IDFT (DFT). A comb with $Q$ comb lines thus supports an equal number of ODDM channels. A delay line of length $L_0$ simply circularly increments the indices of the coefficients $\widetilde{v}_n$ by one. This mapping between symbols and the encoded data is similar to OFDM, but the result of the data mapping is applied to comb line amplitudes instead of time domain samples. Simply put, the role of time and frequency have been exchanged compared to OFDM.

The analysis can be expanded to a more general class of carriers, including low-coherence continuous wave (CW) optical carriers as initially assumed in Section 2. Generally, the ODDM network relies on delayed versions of the light source being orthogonal with each other and providing separate carriers, i.e.,

$$\frac{1}{T_{UI}}\int_0^{T_{UI}} E_c(t - n\tau_0)E_c^*(t - n'\tau_0)\, dt = P_c\delta_{n,n'} \tag{14}$$

which was shown above for a square shaped comb source. Generally, for $N$ such carriers to be orthogonal with each other, an equal number of degrees of freedom needs to be available. In the case of a comb source, these are given by the number of comb lines, i.e., at least $N$ comb lines need to be present.

The analysis for a low coherence CW source starts from a different perspective in Section 2, in which sufficiently large differential delays ensure the absence of interference. The introduced criterion was $L_0 > L_c$, i.e., the delay length has to be larger than the coherence length of the light. Strictly, this only ensures that $E_0(t - n\tau_0)$ and $E_0(t - n'\tau_0)$ have uncorrelated, random phases. For Eq. (14) to hold, the integral also has to be integrated over a sufficiently long time-duration such that these random phases create a close to zero average. For a pair of two carriers, this is verified if the UI is substantially larger than the coherence time of the light, $\tau_c = 1/\Delta\nu$. This integration time has to be further increased when a large number of channels have to be mutually orthogonal. Since the phase of $E_c$ remains coherent over a duration $\sim\tau_c$, in order for delayed carriers $E_c(t - n\tau_0)$, $n \in [0, Q-1]$, to form the basis of a $Q$-dimensional vector space, the integration time has to be increased to $\sim Q\tau_c$. The integration time has thus to be increased by a factor $Q$ to maintain a given level of SNR and crosstalk between the ODDM channels. This results in a UI $Q/\Delta\nu$, which is identical to the one assumed for the comb-based system, for which we had $\Delta\nu/Q = \delta\nu$. As an important difference to a comb source, for which Eq. (14) is deterministic, the orthogonality between mutually delayed carriers is stochastic for a low coherence CW source and depends on the evolution of the phase noise for a specific numerical example. Hence, there is an inherent trade-off between the length of the gating periods and the achievable SNR.

**Appendix II: Broadband photonic device design**

Figure 12(a) depicts the optically wideband DCS used in the demodulators, which has been designed into the 220 nm SOI device layer. It includes two 12.4 μm-long coupler sections ($l_{cs}$) with 500 nm-wide waveguides separated by a 200 nm gap ($g_{cs}$), connected by a 6.6 μm-long phase control section ($l_{pcs} + 2l_{tap}$). The latter section includes 1-μm-long tapers ($l_{tap}$) that convert the waveguide widths to 600 nm (top, $w_{top}$) and 400 nm (bottom, $w_{bot}$) and introduce different phase offsets in the two branches. The 90-degree bends on the outside ports of the device have a 5 μm radius. The total device length is 41.4 μm.

Figure 12(b) displays its power transfer characteristics. Since the left and the right parts of the device are symmetrical, the cross-coupling coefficients $S_{41}$ and $S_{32}$ are equal, resulting in identical group delays $\tau_{41} = \tau_{32}$. The power coupling coefficients stay between 0.46 and 0.52 in the 120 nm wavelength range between 1.5 and 1.62 μm. However, the group delays associated with $S_{31}$ and $S_{42}$ differ from each other and are offset by ±9 fs from $\tau_{41/32}$ at 1.55 μm in the device drawn in Fig. 1(a). This is compensated for by small waveguide segments symmetrically added to the right and left output ports of the device, balancing out the four optical paths at 1.55 μm (their locations are indicated by the color-coded red and green labels in the schematic, wherein equal colors indicate compensating waveguide segments of equal length). Nonetheless, group delay mismatches persist at other wavelengths due to the different dispersion, as shown in Fig. 12(c), but remains within ±1 fs in the wavelength range 1.51 to 1.64 μm.

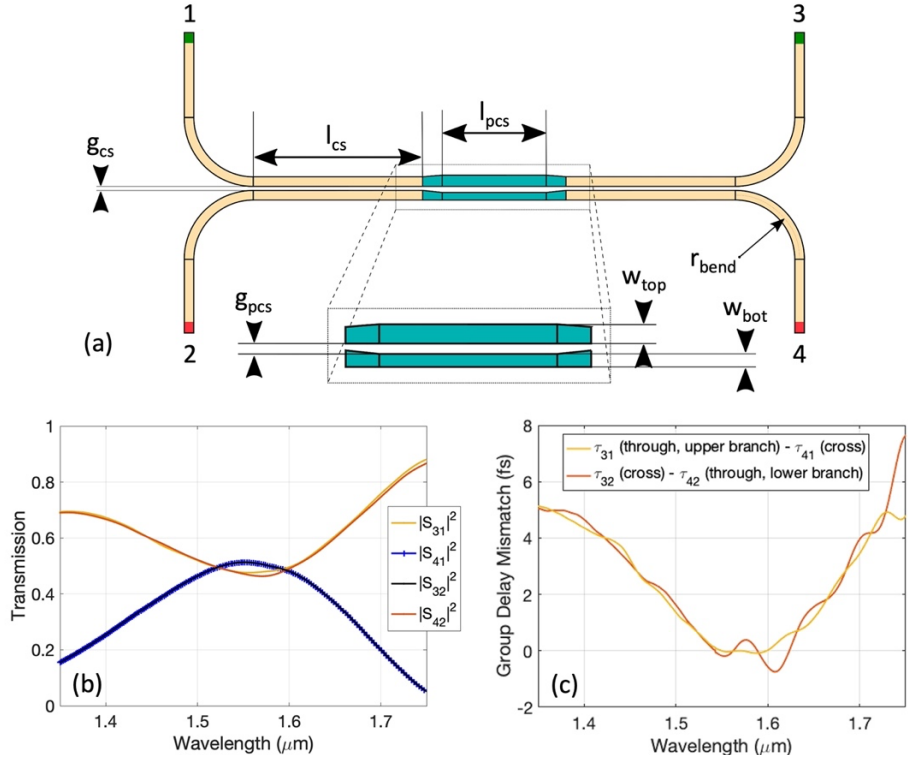

Fig. 12. DCS design and characteristics. (a) layout of the DCS designed into the 220 nm SOI device layer, (b) power transmission coefficients, and (c) group delays associated to the four port combinations labeled in (a) after balancing with external waveguide segments sized for 1550 nm.

Figure 13(a) shows the design of the star coupler in the 400 nm SiN layer of the AMF process. The central slab region is defined by arcs of circles with a radius of $R = 136$ μm and centered at $x = \pm 96$ μm relative to the center point of the device, resulting in a slab width of 80 μm. These centers of curvature are offset by 56 μm relative to the center point of the opposite slab edge to account for diffraction in the waveguide array before coupling into the slab. They are chosen such that the phase front of light injected from the center waveguide of one slab edge follows the opposite slab edge when it reaches it. Waveguides are $w = 500$ nm wide, significantly below the maximum single mode waveguide width, to expand the field profiles to a mode field diameter of 1.2 μm and strongly couple the waveguides to each other at the slab interface,[64,65] where they are spaced by $d = 700$ nm and oriented such that their optical axes cross at the slab edge's center of curvature. The waveguides are located at positions $l \cdot (w + d)$ along the slab edge, with $l$ varying in steps of one. About ~1 dB of the total ILs result from mismatches at the boundaries between the waveguide arrays and the slab, in which the vertical field confinement is higher.

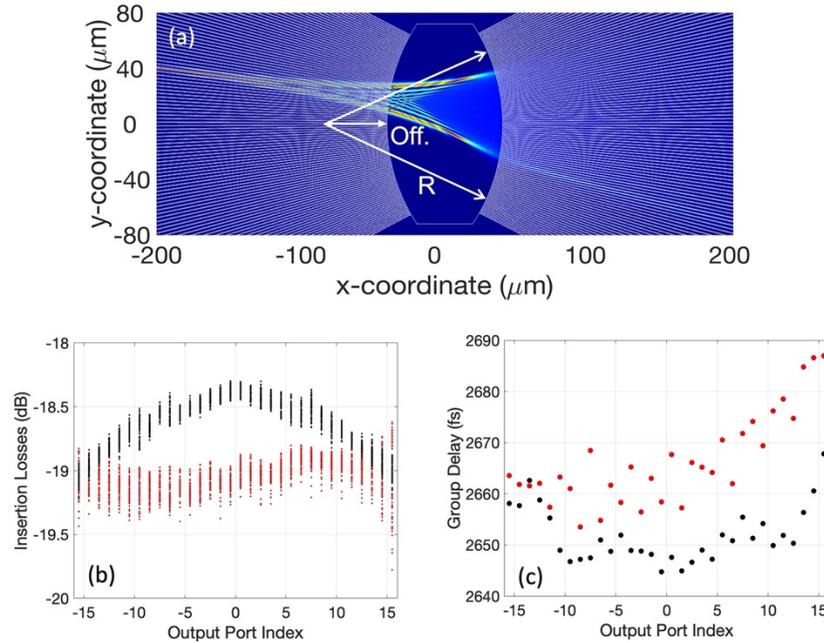

Fig. 13. Star coupler design and characteristics. (a) Layout of the star coupler with overlaid field intensity for light injected through the port of index -15.5. (b) Power transfer coefficients and (c) group delays for light injected through one of the two centermost waveguides (black) and light injected through the waveguide of index $l = -15.5$ (red markers).

Figure 13(b) shows the coupling losses between a central waveguide of the input array and the 32 inner waveguides of the output array (black), as well as between the outer waveguide of index -15.5 and these 32 output waveguides (red) for wavelengths between 1.5 and 1.6 μm. In addition to the 15 dB attenuation resulting from the desired 1/32 splitting, there are less than 5 dB of additional excess losses, for all port combinations. Figure 13(c) shows the group delays for these port combinations at 1.55 μm, with the same color coding. These are all within a ±20 fs span, with the maximum deviation resulting from the increased path length between the input port of index -15.5 and the output port of index +15.5. Increasing the slab size while maintaining the extent of the waveguide array constant reduces the path length mismatches, as shown through simple analytical geometry. However, this also leads to higher excess losses, as diffracting light then covers a growing number of unused outer waveguide ports.

Figure 14 shows the floorplan of the entire system. Light is injected via an edge coupler from the left and transits through the splitter network and the delay lines prior to the signals being modulated by an array of phase modulators at the top of the chip. These can be driven via the RF pad frame on the right of the chip (that has a pitch matched to that of an RF PCB) for an input layer or via a flip-chip integrated RF-electronics chip for a hidden / output layer (see below). The distribution network then operates the signal superpositions and distributes them to the demodulator array, each demodulator filling a row. Their outputs are routed to BPD pairs at the bottom of the chip, with pads marked in red. The blue pads above are auxiliary control pads for electronics, setting the gain and offset of the amplifier and providing limited reconfigurability for the activation function. This layout is based on an extension of a four-channel electronic chip represented as on inset to the right, which has already been designed. It comprises a TIA with a programable nonlinearity implementing the activation function, a track-and-hold stage, a modulator driver and a low pass filter for each AN. Since the array of phase shifters from the demodulators is too large to be connected via a pad frame, a programmable driver chip also needs to be flip chipped onto it to configure the network.

The figure further represents the signal connectivity between two photonic blocks. A second PIC identical to the first one is located below it. The RF-electronics chip straddles the two PICs, receiving photocurrents from the first and driving the modulators of the second.

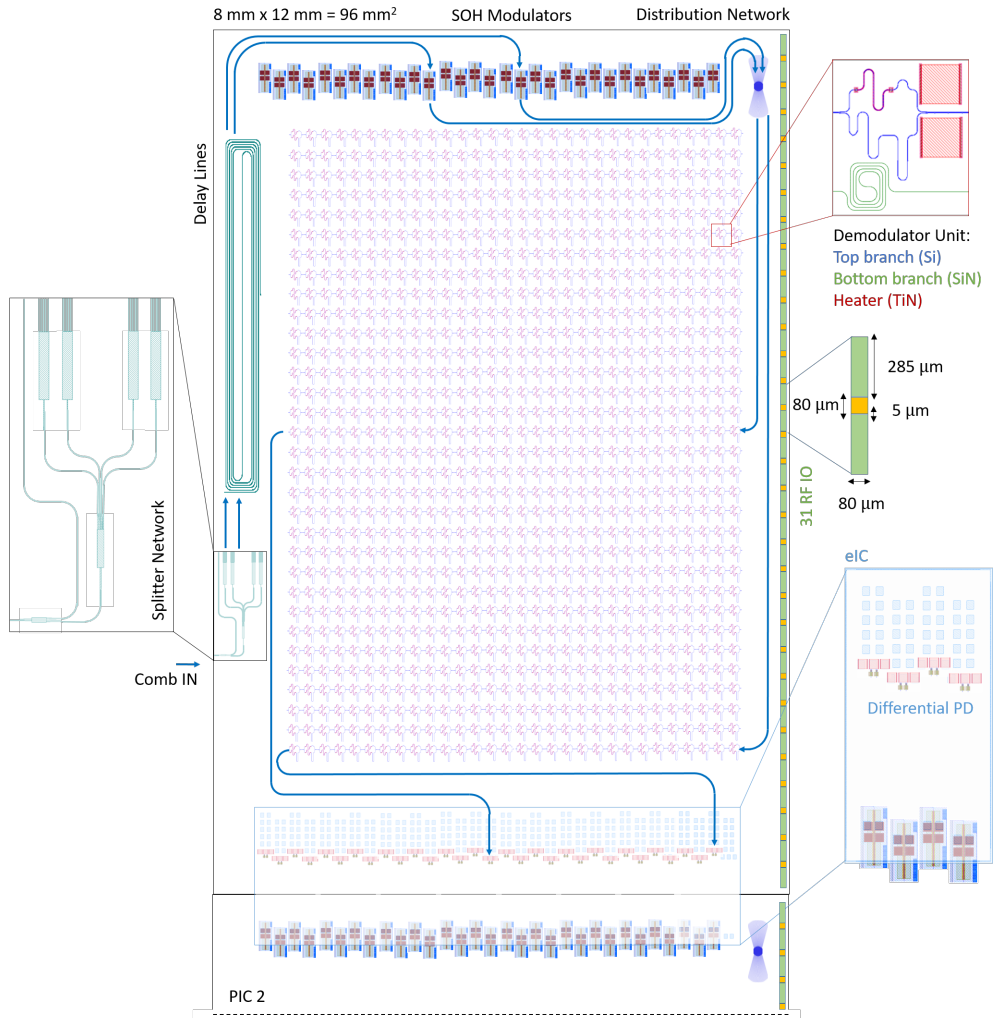

Fig. 14. Floorplan of a 31-by-31 ANN-PIC. The insets show details of the layout.

## Appendix III: Physical Modeling Methodology

The simulation results reported in Section 4 were obtained by modeling a single layer ANN using a dedicated Matlab code.

Light propagation is modeled in the time domain with a time step of 23 fs sufficiently small to model a 350 nm wide optical spectrum with a single sided spectrum (the optical amplitude is represented as a complex valued number with only positive frequencies, a commonly done). 100 symbols with a duration of 20 ps each and sharp transitions are first randomly generated with a uniform distribution between $-\pi$ and $\pi$ for each input neuron, over a total simulation time of 2 ns, prior to being low-pass filtered by a 5$^{th}$ order Bessel filter with identical cutoff frequency as assumed for the electronics, to generate an analog waveform with a commensurate spectral content.

The initial comb spectrum is first generated with an ideal secant squared shaped power spectral distribution, to which random static phases are applied to disperse the comb. To model the optical and RF linewidths, a time series of random phases are generated for both, corresponding to two independent Wiener processes. To model the correlated phase noise corresponding to the optical linewidth, the corresponding phase is applied to the spectrum at each time step prior to applying an inverse Fourier transform (IFT). To model the RF linewidth, the corresponding phase error is first multiplied, for each frequency, by the frequency offset relative to the center of the comb, in units of FSR. This too is done at every time step prior to applying the IFT.[45] RIN and amplified spontaneous emission (ASE) noise are finally applied, RIN by multiplying the amplitude with a random Gaussian distribution of mean 1, lowpass filtered according to the assumed electrical filter bandwidth, and ASE by adding to corresponding noise background to the spectrum. The resulting field is assumed to be injected into the optical input port of the network.

For each Tx network branch, optical losses, dispersion, time delays, and, except for the reference branch, modulation are applied. Dispersion and time delays are applied as frequency dependent phases in the Fourier domain, modulation as multiplication with a phasor in the time domain. The distribution network is modeled as generating linear superpositions of its inputs, with insertion losses and a worst-case differential group delay error extracted from finite-difference time-domain FDTD simulations. After modeling the input stage of the demodulators as a 1-by-2 splitter, group delays and dispersion are applied to the top and bottom branches. The wavelength dependent group delay error resulting from the cascaded DCSs in the top demodulator branch (Fig. 4), as shown in Fig. 12(c), is applied as a worst case, assuming that the light only transits through the top or only through the bottom DCS branches, which results in the largest cumulative error. The final stage of the demodulator is finally modeled as a 2-by-2 3-dB DCS generating a superposition of the fields from the top and bottom demodulator branches in quadrature. The light intensities at the two optical output ports of the demodulator are then converted into a differential photocurrent scaled according to the assumed responsivity. For later computation of the shot noise, the sum of both photocurrents is also recorded in addition to their difference.

In the electrical domain, thermal and shot noise are first modeled in the time domain as white Gaussian noise with the appropriate std. dev. (cf. Section 4.2.), wherein the std. dev. of the shot noise depends on the sum of photocurrents, that is much larger than their difference since both photodiodes generate currents of equal polarity. This photocurrent sum has to be taken, since the shot noise processes associated to the two photocurrents are independent from each other and a sum of variances has thus to be taken when computing the total noise. After applying the noise, in a final step, the differential photo-current is low-pass filtered by a 5$^{th}$ order Bessel filter according to the assumed properties of the electronics.

To generate the graphs as shown in Fig. 5(a), the modeled differential photocurrents are first rescaled to be in a range between -1 and 1, with coefficients obtained from a noiseless simulation to prevent the presence of noise from impacting them. This is then compared to the signal that would have been obtained in the absence of noise and distortion, i.e., the sine of the lowpass filtered analog waveform described above (the signal reference). Since even with an ideal optical network, the generated differential photocurrent is lowpass filtered a second time by the receiver electronics, the signal reference is also filtered a second time by the Bessel filter. This ensures that the result of the comparison reflects the signal integrity as afforded by the optical network and not the interplay between Baud rate and electronic bandwidth, which is a trivial problem and not the object of this evaluation.

**Funding.** German Federal Ministry of Education and Research (BMBF) (03ZU1106BA, 03ZU1106CA). Deutsche Forschungsgemeinschaft (DFG) (403188360).

**Acknowledgements.** Simulations were performed with computing resources granted by RWTH Aachen University under project rwth0938.

**Disclosures.** No conflicts of interest.

**Data availability.** Data underlying the results presented in this paper are not publicly available at this time but may be obtained from the authors upon reasonable request.